\documentclass[aps,prd,onecolumn,groupedaddress,showpacs,nofootinbib,amssymb]{revtex4}
\usepackage[dvips]{graphicx}
\usepackage{amssymb}
\usepackage{amsmath}
\usepackage{graphicx}
\usepackage{amsfonts}
\usepackage{bm}

\begin{document}

\title{Bouncing cosmology with future singularity from  modified gravity}
\author{
S.~D.~Odintsov,$^{1,2,4}$\,\thanks{odintsov@ieec.uab.es}
V.~K.~Oikonomou,$^{3,4}$\,\thanks{v.k.oikonomou1979@gmail.com}}
\affiliation{ $^{1)}$Institut de Ciencies de lEspai (IEEC-CSIC),
Campus UAB, Carrer de Can Magrans, s/n\\
08193 Cerdanyola del Valles, Barcelona, Spain\\
$^{2)}$ ICREA,
Passeig LluA­s Companys, 23,
08010 Barcelona, Spain\\
$^{3)}$ Department of Theoretical Physics, Aristotle University of Thessaloniki,
54124 Thessaloniki, Greece\\
$^{4)}$ National Research Tomsk State University, 634050 Tomsk, Russia and Tomsk State Pedagogical University, 634061 Tomsk, Russia\\
}

\begin{abstract}
We investigate which Jordan frame $F(R)$ gravity can describe a Type IV singular bouncing cosmological evolution, with special emphasis given near the point at which the Type IV singularity occurs. The cosmological bounce is chosen in such a way that the bouncing point coincides exactly with the Type IV singularity point. The stability of the resulting $F(R)$ gravity is examined and in addition, we study the Einstein frame scalar-tensor theory counterpart of the resulting Jordan frame $F(R)$ gravity. Also, by assuming that the Jordan frame metric is chosen in such a way so that, when conformally transformed in the Einstein frame, it yields a quasi de Sitter or de Sitter Friedmann-Robertson-Walker metric, we study the observational indexes which turn out to be consistent with Planck 2015 data in the case of the Einstein
frame scalar theory. Finally, we study the behavior of the effective equation of state corresponding to the Type IV singular bounce and after we compare the resulting picture with other bouncing cosmologies, we critically discuss the implications of our analysis.
\end{abstract}


\maketitle



\def\pp{{\, \mid \hskip -1.5mm =}}
\def\cL{\mathcal{L}}
\def\be{\begin{equation}}
\def\ee{\end{equation}}
\def\bea{\begin{eqnarray}}
\def\eea{\end{eqnarray}}
\def\tr{\mathrm{tr}\, }
\def\nn{\nonumber \\}
\def\e{\mathrm{e}}

\section{Introduction}

Bouncing cosmology \cite{bounce1,bounce2,bounce3,bounce4,bounce5,quintombounce,ekpyr1,ekpyr2,ekpyr3,superbounce2} provides us with an appealing solution of the initial singularity problem, which is a rather unwanted feature in cosmological theories. This is because, in the context of bouncing cosmology, the Universe contracts until a minimal radius is reached, and after that point it expands. Therefore, the Universe never collapses to a singular point, thus avoiding the initial singularity. The cosmological bounces can appear in two main categories, the Loop Quantum Cosmology \cite{LQC} matter bounce theories \cite{matterbounce} and also in theories that make use of scalar fields in order to avoid singularities and generate bounces \cite{bounce1,bounce2,bounce3,bounce4,bounce5,quintombounce,ekpyr1,ekpyr2,ekpyr3,matterbounce}. In addition to these, modified gravity also offers a consistent description of bouncing cosmology \cite{superbounce2,superbounce3}. In the context of bouncing cosmology, the acceleration and thermal history of our Universe are consistently described (see Refs. \cite{bounce1,bounce2} for review on this issue) and also certain CMB anomalies on large angular scale find a satisfactory explanation \cite{piao}. In order for a bounce to occur in standard Einstein-Hilbert gravity, the null energy condition has to be violated, something that is only possible for physical systems for which the Hamiltonian is bounded from below \cite{vikman2}. Some drawbacks that come along with the appealing features of the bounces are that, bouncing cosmologies suffer from ghost and primordial instabilities during the contracting phase, rendering the contracting phase a problematic era during the evolution. The issue of the contracting phase is known as BKL instability \cite{bkl}. Ghost instabilities can be resolved in the context of Galileon and ghost condensate models \cite{vikman1,bounce3}, and the issue of the contracting phase is successfully resolved in the context of the ekpyrotic contraction theories \cite{bounce4,bounce5,refcai}. Actually, as was shown in Ref. \cite{refcai}, the BKL instability issue can be successfully resolved in the context of an ekpyrotic contraction.

On the other hand, the initial singularity is not the only type of singularity that may occur in a cosmological theory, since there exist other types of milder singularities, which are called sudden \cite{barrowsing1,barrowsing2,barrowsing3,barrow,Barrow:2015ora} or finite time singularities \cite{Nojiri:2005sx,sergnoj}. The initial singularity is a crushing type singularity, in which the strong energy theorems \cite{hawkingpenrose} apply and geodesics incompleteness occurs at these points. However, in the context of finite time singularities, only the Big Rip \cite{Caldwell:2003vq,ref5} is of crushing type, with the rest of the finite time singularities being milder, and also geodesics incompleteness does not necessarily occur for the non-crushing types singularities. In the context of general relativity, crushing type singularities occur during the process of gravitational collapse, and several conjectures have been proposed that point out the need to protect the rest of the Universe from these naked singularities with a ``cloth". This is the cosmic censorship hypothesis \cite{penrosecch}, which up to date has not been proved yet. For an informative account on this issue see \cite{Virbhadra:1995iy,Virbhadra:1998kd,Virbhadra:2002ju}. In cosmology however there is no way to ``dress'' these singular points, therefore the complete understanding of their nature and implications is compelling. It seems that the non-crushing type singularities offer a good testing ground for the complete understanding of finite time singularities and their consequences in the cosmological evolution of our Universe. In Refs. \cite{noo1,noo2,noo3}, we studied the Type IV finite time singularities and their implications in the cosmological evolution (especially, after inflation), in the context of single scalar \cite{noo1}, or multiple scalar fields \cite{noo2,noo3}. In this paper we shall study how a Type IV singular bouncing cosmology can be generated from a general pure $F(R)$ \cite{reviews,importantpapers,Nojiri:2006gh,Capozziello:2006dj,Nojiri:2006be,sergeibabmba} gravity, with pure meaning that no matter fluids are assumed to be present. An important assumption we shall make is that the bouncing point coincides with the point that the Type IV singularity occurs. We shall be particularly interested in finding the $F(R)$ theory which generates the bounce near the bouncing point, which is also the point at which the Type IV singularity occurs. For a list of review and important papers on $F(R)$ theories of gravity, the reader is referred to \cite{reviews,importantpapers,sergeibabmba}. In order to reveal which Jordan frame $F(R)$ gravity can successfully describe the singular bounce near the Type IV singularity, we shall use some very well known reconstruction techniques \cite{Nojiri:2006gh,Capozziello:2006dj,Nojiri:2006be}, and as we demonstrate, the resulting Jordan frame $F(R)$ gravity is an $R^2$ gravity plus cosmological constant. Having found this form of the $F(R)$ gravity, we also investigate the Einstein frame implications of this $F(R)$ gravity, assuming a quasi de Sitter solution in the Einstein frame. The observational implications of the Einstein frame canonical scalar theory are also studied in detail. Finally, we also study the behavior of the effective equation of state (EoS) corresponding to the Type IV singular bounce solution and compare the resulting picture with other bouncing cosmologies.

This paper is organized as follows: In section II we present in brief all the essential information with regards to finite time singularities and also we discuss our motivation to use a Type IV singularity for our study. In section III we study the general properties of the Type IV singular bounce we shall use in the forthcoming sections and in section IV, using well known reconstruction techniques, we investigate which $F(R)$ gravity can successfully describe the Type IV singularity near the singularity point, which we chose to coincide with the bouncing point. Also the stability of the resulting solution is discussed in the end of the section. In section V, we study the Einstein frame canonical scalar-tensor theory corresponding to the Jordan frame $F(R)$ gravity we found in section IV. In addition, we also investigate the observational indices of the Einstein frame scalar theory, assuming a quasi de Sitter solution or a de Sitter solution exists in the Einstein frame. In section VI, we perform a full analysis of the EoS corresponding to the Type IV singular bounce under study, and also we compare the behavior of the Type IV singular bounce, to other bouncing cosmologies. Finally, a critical discussion on the results along with the concluding remarks follow in section VII.

\subsubsection*{Geometric background conventions}

Before we start, it is worth mentioning the background geometric
conventions we shall use in this article. We initially work in
the Jordan frame for all $F(R)$ gravities we shall discuss, and we adopt the metric formalism approach \cite{reviews}. In
addition, we assume that the background geometry consists of a
pseudo-Riemannian manifold, which locally is a Lorentz metric, a
flat FRW one, with line element,
\begin{equation}\label{metricformfrwhjkh}
\mathrm{d}s^2=-\mathrm{d}t^2+a^2(t)\sum_i\mathrm{d}x_i^2
\end{equation}
In this geometric background, the Ricci scalar reads,
\begin{equation}\label{ricciscal}
R=6(2H^2+\dot{H}),
\end{equation}
with $H(t)$ denoting as usual the Hubble rate, and the ``dot''
denotes differentiation with respect to the cosmic time $t$.
Finally, we choose the affine connection on this manifold to be the
Levi-Civita one, which is a metric compatible, symmetric and
torsion-less.

\section{Finite-time singularities essentials}

The classification of finite-time cosmological singularities was
extensively done in a formal way in
Refs.~\cite{Nojiri:2005sx,sergnoj}, and we briefly recall the basic
features of this classification, adopting the notation of
Refs.~\cite{Nojiri:2005sx,sergnoj}. There are four types of finite
time cosmological singularities, the Type I,II,III and Type IV
singularities, and these are classified in the following way
\cite{Nojiri:2005sx,sergnoj},
\begin{itemize}
\item Type I (``Known as Big Rip Singularity'') : This type of
cosmological singularity is the most severe among finite time
cosmological singularities and it is a singularity of crushing type.
It occurs when as the cosmic time approaches a specific time ($t_s$), that
is when $t \to t_s$, the effective energy density
$\rho_{\mathrm{eff}}$, the scale factor $a(t)$, and also the
effective pressure $p_\mathrm{eff}$ diverge, that is, $a \to
\infty$, $\rho_\mathrm{eff} \to \infty$, and
$\left|p_\mathrm{eff}\right| \to \infty$. For an important stream of
papers with regards to the Big Rip singularity, the reader is
referred to, Ref.~\cite{Caldwell:2003vq,Nojiri:2005sx,ref5}
\item Type II (Known as ``Sudden Singularity'') \cite{barrowsing2,barrow}: This singularity occurs when, as the cosmic time approaches $t \to
t_s$, only the scale factor $a$ and the effective energy density
$\rho_{\mathrm{eff}}$ take bounded values, that is, $a \to a_s$,
$\rho_{\mathrm{eff}} \to \rho_s$, with both $a_s,\rho_s<\infty$,
with the effective pressure diverging as $t\to t_s$, that is,
$\left|p_\mathrm{eff}\right| \to \infty$. This case for example
occurs when the second and higher derivatives of the scale factor diverge.
\item Type III : This singularity occurs when, as the cosmic time
approaches $t \to t_s$, only the scale factor remains finite $a \to
a_s$, but both the effective energy density and the corresponding
effective pressure diverge, that is, $\left|p_\mathrm{eff}\right|
\to \infty$ and $\rho_\mathrm{eff} \to \infty$. Equivalently, this
means that the first and higher derivatives of the scale factor diverge.
\item Type IV : This type of singularity is the most mild among all the three aforementioned types of finite time singularities, and we shall focus on
this type of singularity in the following. For a detailed study on
this finite time singularity, see Ref. \cite{Nojiri:2005sx}. This
singularity occurs when, as the cosmic time approaches $t_s$, all
the cosmological physical quantities remain finite, that is, the
effective energy density $\rho_\mathrm{eff} \to \rho_s$, the
effective pressure $\left|p_\mathrm{eff}\right| \to p_s$ and the
scale factor $a \to a_s$, but the higher derivatives of the Hubble
rate diverge.
\end{itemize}
In the next section we shall further analyze in brief our motivation
to use the Type IV kind of finite time cosmological singularity for
our analysis of the singular bounce.

\subsection{Why choosing a Type IV singularity: Brief discussion}

As we already mentioned, among all the types of finite singularities
which we presented previously, the most mild is the Type IV, with
mild referring to the geodesics incompleteness issue that might
occur at finite time singularities. The crushing type singularities,
such as the Big Rip \cite{Caldwell:2003vq} or the initial
singularity \cite{hawkingpenrose}, always lead to severe phenomena,
which when considered classically, these are flaws of the theory,
like the singularity in the Coulomb potential in classical
electrodynamics. For these singularities, the energy theorems are
violated, therefore these can be seen as either flaws of the theory
or indicators that a new, quantum maybe, theory describes the
physical system at the energies that these correspond.

On the other hand, the Type IV singularity is less harmful since no
geodesic incompleteness necessarily occurs and also all the energy
theorems of Hawking and Penrose \cite{hawkingpenrose} are satisfied, so these
singularities are not so severe. However, their presence and
possible consequences should be scrutinized in order to fully
understand what these singularities indicate. In this paper we shall
study a bouncing cosmology that has in its evolution a Type IV
singularity. Notice that in bouncing cosmology no initial
singularity appears, so the only singularity that occurs during the bounce evolution is the Type
IV. These appealing properties of the Type IV singularity motivated
us to realize a non-singular bouncing cosmology by using a Jordan
frame $F(R)$ gravity.

\section{Detailed description of the bounce solution}

In this section we shall present the Type IV singular bouncing
cosmology that we shall extensively study in the following sections.
Before we start, it is worth recalling the basic properties of the
bouncing cosmologies. For details on this see Refs.
\cite{bounce1,bounce2,bounce3,bounce4,bounce5,quintombounce}.

A cosmological bounce consists from two eras, a contraction era and
an expansion era. At the beginning, and during the contraction, the
scale factor decreases, that is $\dot{a}<0$, until the Universe
reaches a minimal radius, where $\dot{a}=0$, it bounces off and
starts to expand, with $\dot{a}>0$.  The fact that the Universe
reaches a minimal radius is what renders the bouncing cosmology so
appealing, since the initial singularity is avoided in this way and
there exists also the possibility of describing successfully early
time acceleration \cite{matterbounce,ekpyr1,ekpyr2,ekpyr3}, thus
avoiding the standard description with inflationary models.

Let us assume that the bouncing point occurs at a cosmic time $t_s$.
When the Hubble rate is taken into account, in the contracting
phase, which means that $t<t_s$ the Hubble rate is negative
$H(t)<0$, at the bouncing point becomes equal to zero , $H(t_s)=0$,
while for $t>t_s$, the Hubble rate is positive, $H(t)>0$.

The singular bounce we shall consider in this paper, has the
following scale factor,
\begin{equation}\label{scalebounce}
a(t)=e^{f_0\left(t-t_s\right)^{2(1+\varepsilon)}},
\end{equation}
where we normalized the scale factor to be equal to one at the
bouncing point, that is $a(t_s)=1$, and also $\varepsilon$ and $f_0$
are constant parameters, with the values that $\varepsilon$ is
allowed to take to be specified shortly. The Hubble rate
corresponding to the scale factor (\ref{scalebounce}) is equal to,
\begin{equation}\label{hubblebounce}
H(t)=2 (1+\varepsilon ) f_0 \left(t-t_s\right)^{2\varepsilon+1 }.
\end{equation}
Note that in the Planck unit system, the Hubble rate is measured in $e\mathrm{V}$, the time is measured in $(e\mathrm{V})^{-1}$, so the parameter $f_0$ is measured in $(e\mathrm{V})^{2\varepsilon+2}$. However, we shall express time in seconds, so the Hubble rate is measured in $(\mathrm{sec})^{-1}$ and consequently the parameter $f_0$ is measured in $(\mathrm{sec})^{-2\varepsilon-2}$. We adopt these units in the rest of this paper. 

Taking into account the classification of singularities we presented
in the previous section, the Type IV singularity occurs when the
exponent in Eq. (\ref{hubblebounce}), is $2\varepsilon+1>1$, which
means that $\varepsilon >0$, so for all positive values of the
parameter $\varepsilon$. However, we shall assume
that $\varepsilon < 1$ in which case, the bounce
(\ref{scalebounce}) is a small deformation of the well known
\cite{sergeibabmba} bouncing cosmology,
\begin{equation}\label{bambabounce}
a(t)=e^{f_0\left(t-t_s\right)^{2}}.
\end{equation}
Notice that the Type IV singularity
occurs at $t=t_s$, which is exactly the bouncing point.

It is worth elaborating on the Type IV singularity caused by the structure of the Hubble rate (\ref{hubblebounce}). Set for simplicity $2\varepsilon+1=\beta $, so that the Hubble rate becomes,
 \begin{equation}\label{hubblebounceref}
H(t)=2 (1+\varepsilon ) f_0 \left(t-t_s\right)^{\beta }.
\end{equation}
Let us see for which values of $\beta $ the Type IV occurs. In the list below we quote all the possibilities for a finite singularity to occur, for various values of the parameter $\beta$,
\begin{itemize}\label{lista}
\item $\beta<-1$ corresponds to the Type I singularity.
\item $-1<\beta<0$ corresponds to Type III singularity.
\item $0<\beta<1$ corresponds to Type II singularity.
\item $\beta>1$ corresponds to Type IV singularity.
\end{itemize}
The Type IV singularity case occurs when $\beta>1$. This can be easily seen, since in the Type IV case, the singularity occurs when the higher derivatives of the Hubble rate are divergent, which means that,
\begin{equation}\label{refhubcond}
\frac{\mathrm{d}^nH(t)}{\mathrm{d}t^n}\rightarrow \infty\, ,
\end{equation}
for some $n\geq 2$. Let us compute the lowest derivative for which the Type IV singularity could occur, which is for $n=2$, 
\begin{equation}\label{presenth}
\frac{\mathrm{d}^2H(t)}{\mathrm{d}t^2}=2 (1+\varepsilon ) f_0\beta (\beta-1)\left(t-t_s\right)^{\beta-2 }\, ,
\end{equation}
which is clearly divergent when $1<\beta<2$. If $\beta >2$, then the second derivative of the Hubble rate is finite, but the third derivative becomes divergent,
\begin{equation}\label{presenth1}
\frac{\mathrm{d}^3H(t)}{\mathrm{d}t^3}=2 (1+\varepsilon ) f_0\beta (\beta-1)(\beta-2)\left(t-t_s\right)^{\beta-3 }\, .
\end{equation}
In order for the bounce (\ref{scalebounce}) to be a deformation of the bounce (\ref{bambabounce}), we assume that $\varepsilon < 1$. Since $\beta=2\varepsilon+1$, this means that certainly $\beta>1$ but also that $\beta<2$, so $1<\beta<2$, which means that the second derivative of the Hubble rate (\ref{hubblebounce}), is divergent (see Eq. (\ref{presenth})). Consequently a Type IV singularity occurs for $0<\varepsilon< \frac{1}{2}$ and for these values of $\varepsilon$, the bounce (\ref{scalebounce}) is a deformation of the bounce (\ref{bambabounce}), since $\varepsilon <1$. We have to note that we also restrict ourselves to positive values of the parameter $\varepsilon$.

Before we proceed, we need to address another issue where a possible inconsistency might occur. Throughout the article we shall assume that $0<\varepsilon< \frac{1}{2}$ and the values that $\varepsilon$ is allowed to take are chosen is such a way so that the scale factor and the Hubble rate never become complex. With regards to the scale factor, this would require that the exponent of $t-t_s$ in Eq. (\ref{scalebounce}), takes the following form,
\begin{equation}\label{expoenentsca}
(t-t_s)^{2(\varepsilon+1)}=(t-t_s)^{\frac{2n}{2m+1}}\, ,
\end{equation}
with $n$ and $m$ being arbitrary integers appropriately chosen so that $0<\varepsilon <\frac{1}{2}$. One convenient choice that never makes the scale factor complex but also renders the parameter $\varepsilon$ smaller than one, is for $n=12$ and $m=5$. In this case, the scale factor (\ref{scalebounce}) reads,
\begin{equation}\label{scalebounceref1}
a(t)=e^{f_0\left(t-t_s\right)^{\frac{24}{11}}}=e^{f_0\left(\left(t-t_s\right)^{24}\right)^{\frac{1}{11}}}\, ,
\end{equation}
which is never complex, for all $t$\footnote{Note that if we take the scale factor to be equal to $a(t)=e^{f_0\left(\left(t-t_s\right)^{\frac{1}{11}}\right)^{24}}$, then for $t<t_s$, the expression of the scale factor contains $(-1)^{\frac{1}{11}}$, which has one negative but real branch and two complex branches, that is $(-1)^{\frac{1}{11}}=-1,0.959493 + 0.281733 i,0.959493 - 0.281733 i$, so by keeping the real branch, we obtain again a real number, since the resulting scale factor contains $(-1)^{24}$, which is positive}. For the choice of $n$ and $m$, the parameter $\varepsilon$ reads $\varepsilon=\frac{1}{11}$, so we use this value hereafter. For $\varepsilon=\frac{1}{11}$, the Hubble rate reads,
\begin{equation}\label{hubblebounceref1ere}
H(t)=2 (1+\varepsilon ) f_0 \left(t-t_s\right)^{2\varepsilon+1 }=2 (1+\varepsilon ) f_0 \left(t-t_s\right)^{2\varepsilon }\left(t-t_s\right)=\left(t-t_s\right)=2 (1+\varepsilon ) f_0 \left(t-t_s\right)^{\frac{22}{11} }\left(t-t_s\right)\, ,
\end{equation}
which clearly never takes complex values (for the same reason as in the scale factor case, since $(t-t_s)^{\frac{24}{11}}=\left((t-t_s)^{24}\right)^{\frac{1}{11}}$), but can become negative for $t<t_s$ due to the last term $(t-t_s)$. Having cleared out this vague spot, let us see now how the scale factor and the Hubble rate behave as
functions of cosmic time. To this end, we choose for illustrative
purposes only, $t_s=10^{-35}$sec, $\varepsilon=\frac{1}{11}$ and
$f_0=0.001(\mathrm{sec})^{-2\varepsilon-2}$.
\begin{figure}[h] \centering
\includegraphics[width=15pc]{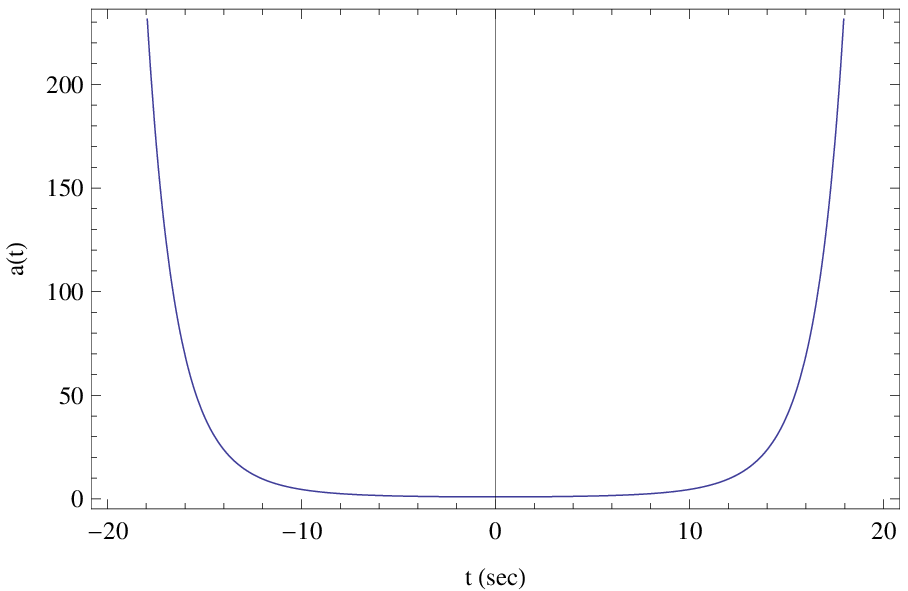}
\includegraphics[width=15pc]{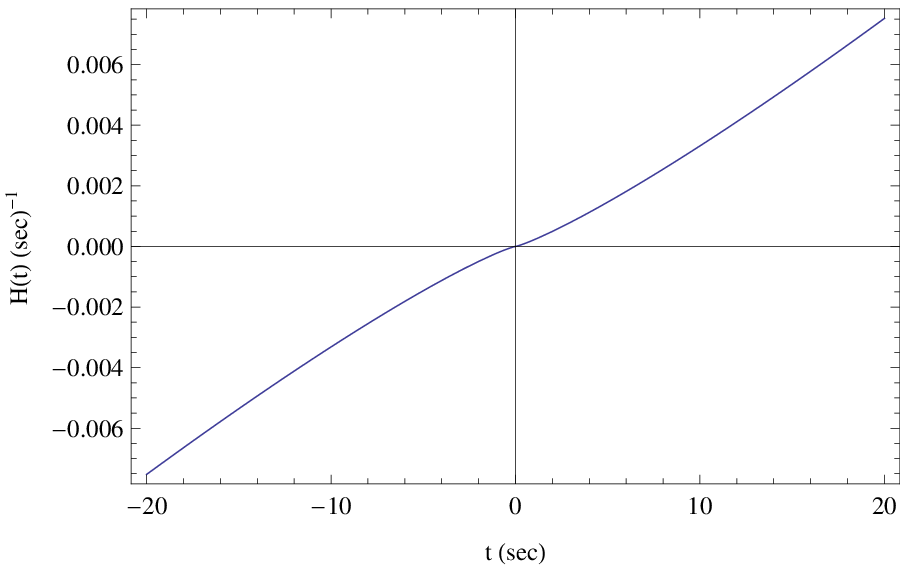}
\caption{The scale factor $a(t)$ (left plot) and the Hubble rate
(right plot) as a function of the cosmic time $t$, for
$t_s=10^{-35}$sec, $\varepsilon=\frac{1}{11}$ and $f_0=0.001(\mathrm{sec})^{-2\varepsilon-2}$, for $a(t)=e^{f_0\left(t-t_s\right)^{2(1+\varepsilon)}}$}
\label{plot1}
\end{figure}
In Fig. \ref{plot1}, we plot the time dependence of both the scale
factor (\ref{scalebounce}) and of the Hubble rate
(\ref{hubblebounce}). As we can see all the qualitative features of
the bounce are satisfied, that is, before the bounce $H(t)<0$, at
the bounce $H(t)=0$, and after the bounce $H(t)>0$. In addition, the
contraction and expansion can be observed by looking the behavior of
the scale factor, since for $t<t_s$, the scale factor decreases
until a minimal value is reached at $t=t_s$, and then for $t>t_s$
the Universe expands.
\begin{figure}[h] \centering
\includegraphics[width=15pc]{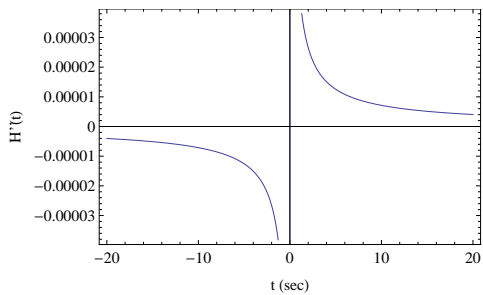}
\caption{The second derivative of the Hubble rate $H''(t)$ as a function of the cosmic time $t$, for
$t_s=10^{-35}$sec, $\varepsilon=\frac{1}{11}$ and $f_0=0.001(\mathrm{sec})^{-2\varepsilon-2}$, for the bounce $a(t)=e^{f_0\left(t-t_s\right)^{2(1+\varepsilon)}}$}
\label{newplotref}
\end{figure}
Finally, in Fig. \ref{newplotref}, we plot the behavior of the second derivative of the Hubble rate $H''(t)$ as a function of time, for
$t_s=10^{-35}$sec, $\varepsilon=\frac{1}{11}$ and $f_0=0.001(\mathrm{sec})^{-2\varepsilon-2}$. As we can see, at the Type IV singularity, the function $H''(t)$ blows-up for $t\rightarrow t_s$, as expected.

\section{Bounce solution from Jordan frame $F(R)$ gravity }

The $F(R)$ modified gravity theoretical framework can realize cosmological scenarios, that were, to some extent, exotic for ordinary general
relativity. Noteworthy is the fact that reconstructed $F(R)$ gravity models can have a direct relation with viable gravity models based on the Khoury chameleon scenario \cite{khoury}. In this section, by using well known
reconstruction techniques
\cite{Nojiri:2006gh,Capozziello:2006dj,Nojiri:2006be},
we shall investigate which $F(R)$ gravity can generate the Type IV
singular bouncing cosmology of Eq. (\ref{scalebounce}). The focus
will be given for times near the Type IV singularity, that is when
$t\rightarrow t_s$. For important reviews and
papers on the reconstruction techniques we shall use, consult
\cite{reviews,importantpapers} and references therein.

Consider the following Jordan frame $F(R)$ gravity, with action,
\begin{equation}
\label{action} \mathcal{S}=\frac{1}{2\kappa^2}\int
\mathrm{d}^4x\sqrt{-g}F(R)+S_m\, ,
\end{equation}
where $\kappa$ is related to Newton's constant $\kappa^2=8\pi G$,
and in addition $S_m$ denotes the action of all matter fluids
present. In the context of the metric formalism of $F(R)$ gravity
\cite{reviews}, upon variation with respect to the metric $g_{\mu
\nu}$, we obtain the following equations of motion,
\begin{align}
\label{modifiedeinsteineqns}
R_{\mu \nu}-\frac{1}{2}Rg_{\mu \nu}=\frac{\kappa^2}{F'(R)}\left( T_{\mu
\nu}+\frac{1}{\kappa^2} \left(\frac{F(R)-RF'(R)}{2}g_{\mu \nu}+\nabla_{\mu}\nabla_{\nu}F'(R)-g_{\mu
\nu}\square F'(R) \right) \right) \, .
\end{align}
where the prime denotes differentiation with respect to the
curvature scalar $R$ and as usual, $T_{\mu \nu}$ is the energy
momentum tensor that receives contributions from all matter fluids.
By observing Eq. (\ref{modifiedeinsteineqns}), it is obvious that
the energy momentum tensor receives an extra contribution which
originates from the $F(R)$ gravitational sector. This extra
contribution, sometimes called geometric contribution, is what makes
$F(R)$ gravity a modified theory of standard Einstein-Hilbert
gravity. Particularly, the extra contribution to the energy momentum
tensor is equal to,
\begin{equation}
\label{newenrgymom}
T^\mathrm{eff}_{\mu \nu}=\frac{1}{\kappa^2}\left( \frac{F(R)-RF'(R)}{2}g_{\mu
\nu}+\nabla_{\mu}\nabla_{\nu}F'(R)-g_{\mu \nu}\square F'(R)\right)\, .
\end{equation}
In addition, we assume that the spacetime metric is the flat FRW of
Eq. (\ref{metricformfrwhjkh}).

What we are interested in now, is to find which pure $F(R)$ gravity
can generate the cosmological evolution of the singular bounce with
scale factor given in Eq. (\ref{scalebounce}). With pure $F(R)$ it
is meant that no matter fluids are assumed to be present. We make use of quite well known reconstruction techniques
\cite{Nojiri:2006gh,Capozziello:2006dj,Nojiri:2006be} and we shall
focus our study for cosmological times near the Type IV singularity,
which occurs at $t=t_s$. Start from a general pure $F(R)$ gravity in
the Jordan frame with action,
\begin{equation}
\label{action1dse}
\mathcal{S}=\frac{1}{2\kappa^2}\int \mathrm{d}^4x\sqrt{-g}F(R)\, .
\end{equation}
The first FRW equation can be obtained upon variation of action
(\ref{action1dse}), with respect to the metric tensor $g_{\mu \nu}$,
and reads,
\begin{equation}
\label{frwf1}
 -18\left ( 4H(t)^2\dot{H}(t)+H(t)\ddot{H}(t)\right )F''(R)+3\left (H^2(t)+\dot{H}(t) \right )F'(R)-\frac{F(R)}{2}=0\, .
\end{equation}
The reconstruction technique we shall use, involves an auxiliary
scalar field $\phi$, which enters the action of
Eq.~(\ref{action1dse}), in the following way,
\begin{equation}
\label{neweqn123}
S=\int \mathrm{d}^4x\sqrt{-g}\left ( P(\phi )R+Q(\phi ) \right )\, .
\end{equation}
In the absence of a kinetic term for the scalar field, this is
practically a non-dynamical degree of freedom, this is why we called
it an auxiliary degree of freedom. By varying the action of Eq.
(\ref{neweqn123}), with respect to $\phi$, we obtain the following
algebraic relation,
\begin{equation}
\label{auxiliaryeqns}
P'(\phi )R+Q'(\phi )=0\, ,
\end{equation}
with $P'(\phi)=\mathrm{d}P(\phi)/\mathrm{d}\phi$ and
$Q'(\phi)=\mathrm{d}Q(\phi)/\mathrm{d}\phi$. This equation is of
particular importance, since it will provide us with the function
$\phi (R)$, if it can be solved explicitly with respect to the
scalar field. This is a very important point for the reconstruction
method, since by finding $\phi (R)$, we can substitute it in the
following relation,
\begin{equation}
\label{r1} F(\phi( R))= P (\phi (R))R+Q (\phi (R))\, .
\end{equation}
and have explicitly the $F(R)$ gravity. So what we need to know is
the explicit form of $P(\phi )$ and $Q(\phi )$ and in
order to find explicitly their functional form, the action (\ref{neweqn123}) must be
varied with respect to the metric, so we obtain the following
relation,
\begin{align}
\label{r2} 0= & -6H^2P(t)-Q(t )-6H\frac{\mathrm{d}P\left (t\right
)}{\mathrm{d}t}=0\, , \nn 0=& \left ( 4\dot{H}+6H^2 \right )
P(t)+Q(t )+2\frac{\mathrm{d}^2P(t)}{\mathrm
{d}t^2}+4H\frac{\mathrm{d}P(t)}{\mathrm{d}t}=0\, .
\end{align}
Then, by eliminating the function $Q(\phi (t))$ from Eq.~(\ref{r2}),
we get the following second order differential equation,
\begin{equation}
\label{r3} 2\frac{\mathrm{d}^2P(t)}{\mathrm
{d}t^2}-2H(t)\frac{\mathrm{d}P(t)}{\mathrm{d}t}+4\dot{H}P(t)=0\, .
\end{equation}
Notice that in Eqs. (\ref{r2}) and (\ref{r3}), we used the cosmic
time $t$ as a variable of the functions $P(\phi )$ and $Q(\phi )$,
instead of $\phi$. Practically, these variables are identified
within the reconstruction technique we are using, and the relation
$\phi=t$ is valid for a wide range of field values, since the
actions of Eqs. (\ref{action1dse}) and (\ref{neweqn123}), are
equivalent from a mathematical point of view. For details on this
account, see the Appendix of Ref.~\cite{Nojiri:2006gh}. So, for a
known cosmological evolution with a specified Hubble rate, the
solution of the differential equation Eq.~(\ref{r3}), yields the
function $P(t)$, and by substituting the result to Eq.~(\ref{r2}),
we obtain the function $Q(t)$.

Let us proceed to find an analytic form for the function $P(t)$ and
then proceed to find the $F(R)$ gravity near the Type IV
singularity. By using the Hubble rate (\ref{hubblebounce}), the
differential equation (\ref{r3}) reads,
\begin{equation}
\label{ptdiffeqn} 2\frac{\mathrm{d}^2P(t)}{\mathrm {d}t^2}-4
f_0 (1+\varepsilon )\left(t-t_s\right)^{2\varepsilon+1 }
\frac{\mathrm{d}P(t)}{\mathrm{d}t}+8 f_0(1+\varepsilon ) (1+2
\varepsilon ) \left(t-t_s\right)^{2\varepsilon } P(t)=0\, .
\end{equation}
Since we are interested in finding the $F(R)$ gravity near the Type
IV singularity, it is worth changing the variable $t$, to the new
one $x$, defined to be $x=t-t_s$. In this way, as $t$ approaches the
singularity, the variable $x$ approaches zero, or schematically as
$t\rightarrow t_s$, then $x\rightarrow 0$. Using $x$ as a variable,
the differential equation (\ref{ptdiffeqn}) becomes,
\begin{equation}
\label{dgfere} \frac{\mathrm{d}^2P(x)}{\mathrm
{d}x^2}-2f_0(1+\varepsilon
)x^{2\varepsilon+1}\frac{\mathrm{d}P(x)}{\mathrm{d}x}+4 f_0
(1+\varepsilon )(1+2\varepsilon ) x^{2\varepsilon } P(x)=0\, .
\end{equation}
which can be analytically solved by setting $z=x^{2\varepsilon+2}$,
and hence it becomes,
\begin{equation}
\label{diffeqn}
\left(2\varepsilon+2\right)^2z\frac{\mathrm{d}^2P(z)}{\mathrm
{d}z^2}+(2\varepsilon+2)(-2f_0(1+\varepsilon)
z+2\varepsilon+1)\frac{\mathrm{d}P(z)}{\mathrm{d}z}+4 f_0
(1+\varepsilon )(1+2\varepsilon ) P(z)=0\, ,
\end{equation}
The solution to this differential equation is,
\begin{equation}
\label{genrealsol} P(z)= (2 z+2 z \varepsilon )^{\frac{1}{2
(1+\varepsilon )}} C_1 U\left(-\frac{1+4 \varepsilon }{2 (1+\varepsilon )},1+\frac{1}{2 (1+\varepsilon )}, f_0 z
\right) + (2 z+2 z \varepsilon )^{\frac{1}{2 (1+\varepsilon )}} C_2
L_{n}^{m}\left(f_0 z\right)\, ,
\end{equation}
where the functions $U(a,b,z)$ and $L_n^m(z)$ are the confluent
Hypergeometric function and the generalized Laguerre polynomial
respectively, the parameters $C_1,C_2$ are constant arbitrary real
numbers and finally $m$ and $n$ stand for,
\begin{equation}
\label{gfgrg} n=\frac{1+4 \varepsilon }{2 (1+\varepsilon )},{\,}{\,}{\,}m=\frac{1}{2 (1+\varepsilon )}\, ,
\end{equation}
In terms of the variable $x$, the function $P(x)$ reads,
\begin{equation}
\label{genrealsolxfunct} P(x)=(2 x^{2\varepsilon+2}+2
x^{2\varepsilon+2} \varepsilon )^{\frac{1}{2 (1+\varepsilon )}} C_1
U\left(-\frac{1+4 \varepsilon }{2 (1+\varepsilon )},1+\frac{1}{2
(1+\varepsilon )}, f_0 x^{2\varepsilon+2} \right) + (2
x^{2\varepsilon+2}+2 x^{2\varepsilon+2} \varepsilon )^{\frac{1}{2
(1+\varepsilon )}} C_2 L_{n}^{m}\left(f_0 x^{2\varepsilon+2}\right)
\, .
\end{equation}
Then, the function $Q(x)$ easily follows,
\begin{align}
\label{qtanalyticform} Q(x) =& 3\ 2^{1+\frac{1}{2+2 \varepsilon }} f_0 x^{2+2 \varepsilon } (1+\varepsilon )\text{  }(1+\varepsilon )^{\frac{1}{2+2 \varepsilon }}\times \Big{[} -\frac{f_0 (1+4 \varepsilon ) C_1}{1+\varepsilon } U\Big{(} \frac{1-2 \varepsilon }{2+2 \varepsilon },2+\frac{1}{2+2 \varepsilon },f_0 x^{2(1+ \varepsilon )+1} \Big{)})\\ \notag &
+\frac{1}{1+\varepsilon }x^{-2 (1+\varepsilon )} \Big{(}-\Big{(}1+4 f_0 x^{3+4 \varepsilon } (1+\varepsilon )^2\Big{)}) C_1\Big{)} U \Big{(} -\frac{1+4 \varepsilon }{2+2 \varepsilon },1+\frac{1}{2+2 \varepsilon },f_0 x^{2(1+ \varepsilon )+1}\Big{)}
\\ \notag &
+C_2 \Big{(}2 f_0 x^{2+2 \varepsilon } (1+\varepsilon )L_{n_1}^{m_1}\Big{(} f_0 x^{2(1+ \varepsilon )+1 } f_0 \Big{)} \Big{)}  -\left(1+4 f_0 x^{3+4 \varepsilon } (1+\varepsilon )^2\right)L_{n_2}^{m_2}\Big{(} f_0 x^{2(1+ \varepsilon )+1 }\Big{)}\Big{)} \Big{)} \Big{)}\Big{]}\, ,
\end{align}
where we introduced the variables $n_1,m_1$ which are equal to,
\begin{equation}
\label{n1m1} n_1=\frac{-1+2 \varepsilon }{2 (1+\varepsilon )}\,
,\quad m_1=1+\frac{1}{2+2 \varepsilon }\, ,
\end{equation}
and also $n_2$ and $m_2$ stand for,
\begin{equation}
\label{n2ma} n_2=\frac{1+4 \varepsilon }{2+2 \varepsilon }\, , \quad m_2=\frac{1}{2+2 \varepsilon }\, .
\end{equation}
Having at hand the functions $P(x)$ and $Q(x)$, in principle by
substituting in Eq.~(\ref{auxiliaryeqns}), we can obtain the
functional dependence of $x$ as a function of $R$. However, the form
of the functions $P(x)$ and $Q(x)$ does not allow an easy analytic
manipulation of the resulting algebraic equation. However, since we
are interested in the behavior of the $F(R)$ near the bounce, and
since this limiting case is reached in the limit $x\rightarrow 0$,
we shall approximate $P'(x)$ and $Q'(x)$ near the bouncing point. By
taking the derivative of Eq. (\ref{genrealsolxfunct}),
approximating the resulting expression in the limit $x\rightarrow 0$
and by keeping only the leading order terms, the function $P'(x)$
reads,
\begin{equation}
\label{pseries} P'(x)\simeq
\frac{\mathcal{A}}{x^{2(\varepsilon+1)+1}}+\mathcal{O}(x)\, ,
\end{equation}
where we introduced the constant parameter $\mathcal{A}$, which can be found in the Appendix B.

By doing the same to the other function $Q'(x)$, we obtain in the
small $x$ limit,
\begin{equation}\label{qsapprox}
Q'(x)\simeq
\frac{\mathcal{B}}{x^{2(\varepsilon+1)+1}}+\mathcal{C}+\mathcal{O}(x)\,
,
\end{equation}
where the parameters $\mathcal{B}$ and $\mathcal{C}$ can be also found in the Appendix B, and also notice that only leading order terms where kept. Upon substitution of Eqs.~(\ref{pseries}) and (\ref{qsapprox}) into
Eq.~(\ref{auxiliaryeqns}), we obtain the function $x(R)$,
\begin{equation}
\label{finalxr} x\simeq
\left(-\frac{\mathcal{C}}{\mathcal{A}R+\mathcal{B}}\right)^{\frac{1}{2(\varepsilon+1)+1}}\,
.
\end{equation}
Having $x(R)$ at hand, the $F(R)$ gravity near the Type IV
singularity easily follows by making use of Eq.~(\ref{r1}), so the
resulting expression of the $F(R)$ gravity is,
\begin{equation}
\label{finalfrgravity}
F(R)\simeq -\frac{\mathcal{A}^2}{\mathcal{C}}R^2-2\frac{\mathcal{B}\mathcal{A}}{\mathcal{C}}R-\frac{\mathcal{B}^2}{\mathcal{C}}+\mathcal{C}\, .
\end{equation}
Therefore, the $F(R)$ gravity that generates the bounce
(\ref{scalebounce}) near the Type IV singularity is a nearly $R^2$
gravity \cite{starobinsky}. We can bring the resulting expression to
be exactly an Einstein-Hilbert gravity plus curvature corrections by
appropriately choosing the free parameter $C_1$ to satisfy the
following constraint,
\begin{equation}
\label{cond1-B}
-2\frac{\mathcal{B}\mathcal{A}}{\mathcal{C}}=1\, ,
\end{equation}
which holds true if $C_1$ is chosen to be,
\begin{align}\label{c1valueforeinsteihilber}
& C_1=-\frac{(1+2(\varepsilon+1) )^2 \Gamma\Big{(}\frac{2+4
2(\varepsilon+1) }{1+2(\varepsilon+1) }\Big{)}
 \Gamma\Big{(}\frac{3+5 2(\varepsilon+1) }{1+2(\varepsilon+1) }\Big{)}}{12 f_0 (1+3 2(\varepsilon+1) )
 }
 \\ \notag & \times  \frac{1}{\Big{(}(2+2(\varepsilon+1) ) \Gamma\Big{(}\frac{3+5 2(\varepsilon+1) }{1+2(\varepsilon+1) }\Big{)}
 \Gamma\Big{(}1+\frac{1}{1+2(\varepsilon+1) }\Big{)}-2 (1+2 2(\varepsilon+1) ) \Gamma\Big{(}\frac{2+4 2(\varepsilon+1) }{1+2(\varepsilon+1) }\Big{)} \Gamma\Big{(}2+\frac{1}{1+2(\varepsilon+1) }\Big{)}\Big{)}
 \Big{(}\frac{f_0}{1+2(\varepsilon+1) }\Big{)}{}^{-\frac{1}{1+2(\varepsilon+1) }}}\, ,
\end{align}
The resulting $F(R)$ gravity near the Type IV singularity then
reads,
\begin{equation}\label{einsteinlikegrav}
F(R)\simeq R-\frac{\mathcal{A}^2}{\mathcal{C}}R^2-\frac{\mathcal{B}^2}{\mathcal{C}}+\mathcal{C}\, .
\end{equation}
Recall that we assumed
$\varepsilon \ll 1$, and by also assuming that $f_0>0$, it can
easily be shown that the parameter $C_1<0$. Consequently, the
parameter $\mathcal{C}$, defined in Eq. (\ref{mathbandc}), is
negative, therefore the coefficient in front of the $R^2$ in Eq.
(\ref{einsteinlikegrav}) is positive. It is worth redefining the
coefficients of the $F(R)$ gravity, for later convenience, so we
introduce the following new variables,
\begin{equation}\label{coeffcnew}
C_0=-\frac{\mathcal{C}}{4\mathcal{A}^2},{\,}{\,}{\,}\Lambda =-\frac{\mathcal{B}^2}{\mathcal{C}}+\mathcal{C}\, .
\end{equation}
Notice that, since $\mathcal{C}<0$, the coefficient $C_0$ is positive. In
terms of the new coefficients, the Jordan frame $F(R)$ gravity for
the bounce (\ref{scalebounce}), near the Type IV singularity has the
following form,
\begin{equation}\label{jordanframegravity}
F(R)=R+\frac{R^2}{4C_0}+\Lambda \, .
\end{equation}
This Jordan frame $F(R)$ gravity has a particularly interesting
Einstein frame counterpart, since it corresponds to an nearly $R^2$
scalar theory in the Einstein frame. We shall discuss this issue in
detail in a later section.

\subsection{Stability analysis of $F(R)$ gravity solution}

Having at hand the reconstructed $F(R)$ gravity near the bounce, it
is a straightforward task to examine the stability of the solution
near the bounce. What is expected is that the system of differential
equations, when viewed as a dynamical system, is unstable near the
bounce, since the bounce is not an ending state of the system but
just a passing point during the Universe's evolution.

In order to address formally the instability issue, we shall use the
approach adopted in Ref. \cite{Nojiri:2006gh,Capozziello:2006dj,Nojiri:2006be}. The  starting point
of our analysis is the following equation,
\begin{align}\label{staux1}
& 2\frac{\mathrm{d}^2P(\phi)}{\mathrm {d}t^2}-2g'(\phi
)\frac{\mathrm{d}P(\phi )}{\mathrm{d}t}+4g''(\phi ) P(\phi)=0,
\end{align}
where in our case, $g(\phi)=(\phi-t_s)^{2(\varepsilon+1)}$. The
expression in Eq. (\ref{staux1}) can be written as follows,
\begin{align}\label{stauxo1}
&2\frac{\mathrm{d}^2P(\phi )}{\mathrm {d}\phi^2}\Big{(}\frac{\mathrm{d}\phi }{\mathrm {d}t}\Big{)}^2-2\frac{\mathrm{d}P(\phi )}{\mathrm {d}\phi }\frac{\mathrm{d}^2
\phi }{\mathrm {d}t^2}-2g'(\phi )\frac{\mathrm{d}P(\phi )}{\mathrm {d}\phi}\Big{(}\frac{\mathrm{d}\phi }{\mathrm {d}t}\Big{)}^2+4\Big{(}g''(\phi )
\Big{(}\frac{\mathrm{d}\phi }{\mathrm {d}t}\Big{)}^2+g'(\phi )\frac{\mathrm{d}^2\phi }{\mathrm {d}t^2}\Big{)}P(\phi )=0,
\end{align}
and accordingly can be recast as follows,
\begin{align}\label{staux2}
& 2\Big{[}\frac{\mathrm{d}^2P(\phi )}{\mathrm {d}\phi^2}-g'(\phi)\frac{\mathrm{d}P(\phi )}{\mathrm {d}\phi}+g''(\phi )P(\phi )\Big{]}
\Big{(}\Big{(}\frac{\mathrm{d}\phi }{\mathrm {d}t}\Big{)}^2-1\Big{)}
+2\Big{(}\frac{\mathrm{d}P(\phi )}{\mathrm {d}\phi}+2g'(\phi)P(\phi)\Big{)}\frac{\mathrm{d}^2\phi }{\mathrm {d}t^2}=0.
\end{align}
We introduce the function $\delta $ to be equal to,
\begin{equation}\label{staux3}
\delta =\frac{\mathrm{d}\phi }{\mathrm {d}t}-1.
\end{equation}
Practically, the parameter $\delta $ measures the exact way that
perturbations behave for the solutions we presented in the previous
section, with regards to $P(\phi )$, if the system of equations are
treated as a dynamical system. In terms of $\delta $, Eq.
(\ref{staux2}) can be written as follows,
\begin{equation}\label{staux4}
\frac{\mathrm{d}\delta }{\mathrm{d}t}=-\omega (t)\delta,
\end{equation}
with $\omega (t)$ standing for,
\begin{equation}\label{omegarepr}
\omega (t)=2\frac{\frac{\mathrm{d}^2P(\phi )}{\mathrm {d}\phi^2}-g'(\phi)\frac{\mathrm{d}P(\phi )}{\mathrm {d}\phi}+g''(\phi )P(\phi )}
{\frac{\mathrm{d}P(\phi )}{\mathrm {d}\phi}+2g'(\phi )P(\phi )}\Big{|}_{\phi=t}.
\end{equation}
Thereby, if $\omega>0$, the dynamical system of Eq. (\ref{staux4})
is stable, since the coefficient of the dynamical variable $\delta $
is rendered negative. However, if $\omega<0$ the dynamical system is
unstable, since the perturbations grow in an exponential-like way.
Since we are interested in the limit $t\rightarrow t_s$, the
functional form of $\omega (\phi)$, near the Type IV
singularity, where the bounce occurs, reads,
\begin{align}\label{parameterpmega}
& \omega (\phi)\simeq \frac{2 \left(6+\varepsilon
\left(7-\left((\phi-t_s)^2\right)^{\varepsilon } (2+x-4 \varepsilon
)+2 \varepsilon \right)\right)}{(\phi-t_s) \left(-3+\left(-2+4
\left((\phi-t_s)^2\right)^{\varepsilon }\right) \varepsilon
\right)}\,,
\end{align}
which for $\phi\simeq t_s$ can be further approximated by the
following expression,
\begin{equation}\label{lastapprox}
\omega (\phi)\simeq \frac{2 (6+\varepsilon  (7+2 \varepsilon
))}{(\phi-t_s) (-3-2 \varepsilon )},
\end{equation}
As is obvious by looking at Eq. (\ref{lastapprox}), $\omega (\phi)$
is negative for $\phi>t_s$, while it is positive for $\phi<t_s$, so
the system is conditionally unstable (as we anticipated), since the
point $\phi=t_s$ is a saddle point of the dynamical system
(\ref{staux4}).

\section{Einstein frame analysis of the $F(R)$ bounce}

In the previous section we demonstrated that the Jordan frame $F(R)$
gravity of Eq. (\ref{jordanframegravity}) can realize the bounce
cosmology of Eq. (\ref{scalebounce}). In this section we show
that the Einstein frame counterpart of the $F(R)$ gravity
(\ref{jordanframegravity}), is a deformed form of $R^2$ inflation in
the Einstein frame (for a similar approach to ours, see Ref.
\cite{sergeistarobinsky}). Particularly, it corresponds to a
canonical scalar theory with a nearly $R^2$ potential \cite{starobinsky}. This kind of potentials were first studied in \cite{starobinsky}. We have
to note that we do not assume that we start in
the Jordan frame with the scale factor of Eq. (\ref{scalebounce}),
but we start with a convenient metric that, when conformally
transformed to the Einstein frame, it can generate nearly
Starobinsky inflation. We start off with the action of the $F(R)$
gravity (\ref{jordanframegravity}),
\begin{equation}
\label{pure} \mathcal{S}=\frac{1}{2\kappa^2}\int
\mathrm{d}^4x\sqrt{-\hat{g}}F(R)=\frac{1}{2\kappa^2}\int
\mathrm{d}^4x\sqrt{-\hat{g}}\left(R+\frac{R^2}{4C_0}+\Lambda \right)\, ,
\end{equation}
where $\hat{g}_{\mu \nu}$ is the Jordan frame metric tensor. We introduce the auxiliary field $A$, so in terms of this auxiliary
scalar, the action (\ref{pure}) can be written as follows,
\begin{equation}\label{action1dse111}
\mathcal{S}=\frac{1}{2\kappa^2}\int
\mathrm{d}^4x\sqrt{-\hat{g}}\left ( F'(A)(R-A)+F(A) \right )\, .
\end{equation}
Upon variation with respect to the scalar $A$, the solution $A=R$ is
obtained, thus the mathematical equivalence of the actions
(\ref{pure}) and (\ref{action1dse111}) is verified. By using the
canonical transformation,
\begin{equation}\label{can}
\varphi =-\sqrt{\frac{3}{2\kappa^2}}\ln (F'(A)) \, ,
\end{equation}
we are transferred to the Einstein frame. In Eq. (\ref{can}), the
scalar field $\varphi$ is the Einstein frame canonical scalar field.
Conformally transforming the Jordan frame metric $\hat{g}_{\mu \nu
}$,
\begin{equation}\label{conftransmetr}
g_{\mu \nu}=e^{-\varphi }\hat{g}_{\mu \nu } \, ,
\end{equation}
we obtain the Einstein frame scalar field action,
\begin{align}\label{einsteinframeaction}
\mathcal{\tilde{S}}= & \int \mathrm{d}^4x\sqrt{-g}\left (
\frac{R}{2\kappa^2} -\frac{1}{2}\left (\frac{F''(A)}{F'(A)}\right
)^2g^{\mu \nu }\partial_{\mu }A\partial_{\nu }A -\frac{1}{2\kappa^2}
\left ( \frac{A}{F'(A)}-\frac{F(A)}{F'(A)^2}\right ) \right ) \nn =&
\int \mathrm{d}^4x\sqrt{-g}\left ( \frac{R}{2\kappa^2}-\frac{1}{2}g^{\mu
\nu }\partial_{\mu }\varphi\partial_{\nu }\varphi -V(\varphi )\right
)\, ,
\end{align}
with the scalar potential $V(\varphi )$ being equal to,
\begin{align}\label{potentialvsigma}
V(\varphi
)=\frac{A}{F'(A)}-\frac{F(A)}{F'(A)^2}=\frac{1}{2\kappa^2}\left (
e^{\sqrt{2\kappa^2/3}\varphi }R\left (e^{-\sqrt{2\kappa^2/3}\varphi}
\right ) - e^{2\sqrt{2\kappa^2/3}\varphi }F\left [ R\left
(e^{-\sqrt{2\kappa^2/3}\varphi} \right ) \right ]\right ) \, .
\end{align}
For the Jordan frame $F(R)$ gravity (\ref{jordanframegravity}), the
canonical scalar field potential of Eq.~(\ref{potentialvsigma}),
 becomes \cite{sergeistarobinsky},
\begin{equation}\label{vapprox}
V(\varphi)\simeq C_0+C_2e^{-2\sqrt{\frac{2}{3}}\kappa   \varphi
}+C_1e^{-\sqrt{\frac{2}{3}}\kappa   \varphi } \, ,
\end{equation}
where the constant parameters $C_1$ and $C_2$ are related to $C_0$
and $\Lambda$ that appear in Eq. (\ref{jordanframegravity}) as
follows,
\begin{equation}\label{coeffnewdefs}
C_1=-2C_0,{\,}{\,}{\,}C_2=C_0-\Lambda \, .
\end{equation}

The canonical scalar field potential (\ref{vapprox}) corresponds to
a nearly $R^2$ inflationary potential, which in the case that
the coefficient $C_2$ is equal to $C_0$, then the potential becomes
exactly the $R^2$ model potential in the Einstein frame, that
is,
\begin{equation}\label{einsteinfrstarbky}
V(\varphi )=C_0\left (1-e^{\sqrt{\frac{2}{3}}\kappa \varphi}\right
)^2\, .
\end{equation}
This by any means does not imply that the bounce in the Jordan frame
with metric that has a scale factor (\ref{scalebounce}), corresponds
to an inflationary solution when viewed in the Einstein frame. That
would require a very special conformal transformation that may not
necessarily lead to a de Sitter or even quasi de Sitter solution in
the Einstein frame. An interesting scenario occurs when the Jordan frame
$\hat{g}_{\mu \nu}$ is such that, when conformally transformed according to the transformation
(\ref{conftransmetr}), it becomes a de Sitter or quasi de
Sitter metric in the Einstein frame, with the scalar potential being that of Eq. (\ref{vapprox}).
This is a quite appealing scenario and in the next section we shall
consider the observational indices of this nearly $R^2$ model
(\ref{vapprox}). Some interesting studies related to conformal transformation between frames and singularities can be found in \cite{sasaki}.

\subsection{Observational indices of the Einstein frame $F(R)$
gravity}

In this section we shall study the observational implications of the Einstein frame canonical scalar field model with potential that of Eq. (\ref{vapprox}). The potential of Eq. (\ref{vapprox}) in view of the constraints (\ref{coeffnewdefs}), reads,
\begin{equation}\label{skf}
V(\varphi )=C_0+C_2 e^{-2 \sqrt{\frac{2}{3}} \kappa  \varphi }-2 C_0 e^{-\sqrt{\frac{2}{3}} \kappa  \varphi }.
\end{equation}
For the canonical scalar theory in the Einstein frame, assuming a flat FRW metric, the energy density and pressure of the model are equal to,
\begin{equation}\label{ingflatrg}
\rho_{\varphi }=\frac{\dot{\varphi}^2}{2}+V(\varphi ),{\,}{\,}{\,}p_{\varphi }=\frac{\dot{\varphi}^2}{2}-V(\varphi ),
\end{equation}
where as usual, the ''dot'' indicates differentiation with respect to the cosmic time. The Friedmann equations in the presence of the canonical scalar $\varphi $, are given by,
\begin{equation}\label{einsteinfrweqns}
\frac{3H^2}{\kappa^2}=\frac{\dot{\varphi}^2}{2}+V(\varphi ),{\,}{\,}{\,}-\frac{1}{\kappa^2}\left ( 2\dot{H}+3H^2\right )=\frac{\dot{\varphi}^2}{2}-V(\varphi ),
\end{equation}
and also the following second order equation of motion for the scalar field $\varphi $ is satisfied,
\begin{equation}\label{secordereqns}
\ddot{\varphi}+3H\dot{\varphi}=-V'(\varphi ).
\end{equation}
Before getting to the detailed calculation of the observational indices for the model (\ref{skf}), it is worth presenting in brief the essentials of the slow-roll approximation in inflationary theories. For a detailed presentation of these issues, the reader is referred to Refs.~\cite{inflation,inflationreview}. Historically, the slow-roll approximation \cite{linde,steinhard}, was introduced in order to solve the graceful exit of inflation problem and nowadays is frequently used in most inflation predicting theories. The slow-roll conditions are based on the following constraint for the canonical scalar field,
\begin{equation}\label{slowrollconstr}
\frac{1}{2}\dot{\varphi}^2\ll V(\varphi)\, ,
\end{equation}
where it is assumed that this constraint holds true for a sufficiently long period of time, with the latter feature being model dependent nevertheless. The constraint of Eq. (\ref{slowrollconstr}) is called the first slow-roll condition, and it guarantees a long and finite acceleration era. Furthermore, another constraint has to be imposed on the first slow-roll condition, so that the slow-roll accelerating era lasts for a sufficiently long period of time, which is the following, 
\begin{equation}\label{slowrollconstrnew}
|\ddot{\varphi}|\ll \left| \frac{\partial V(\varphi)}{\partial \varphi} \right|\, ,
\end{equation}
The condition of Eq. (\ref{slowrollconstrnew}) is called the second slow-roll condition, which by taking into account the equation of motion of the canonical scalar field, namely Eq. (\ref{secordereqns}), in a flat FRW background, it becomes as follows,
\begin{equation}\label{newcondqweeee}
|\ddot{\varphi}|\ll 3H |\dot{\varphi}| \, .
\end{equation}
In view of the two slow-roll conditions, namely Eqs. (\ref{slowrollconstr}) and (\ref{newcondqweeee}), the equation of motion of the scalar field (\ref{secordereqns}) becomes,
\begin{equation}\label{eqnsmotoinscla}
\dot{\varphi }\simeq -\frac{1}{3H}\frac{\partial V(\varphi )}{\partial \varphi} \, ,
\end{equation}
while the FRW equations of Eq. (\ref{einsteinfrweqns}) read,
\begin{equation}\label{frweqnbecomes}
3 H^2\simeq \kappa^2 V(\varphi ), \,{\,}{\,}{\,} 3H\dot{\varphi }\simeq-V'(\varphi ).
\end{equation}
The two slow-roll conditions, namely Eqs.~(\ref{slowrollconstrnew}) and (\ref{newcondqweeee}), can be shown that they are equivalent to the following two relations,
\begin{equation}\label{sfinalfgwajd}
\left(\frac{V '(\varphi )}{V (\varphi )}\right)^2\ll 2\kappa^2\, , \quad
\left(\frac{V'' (\varphi )}{V (\varphi ^2)}\right)\ll \kappa^2 \, ,
\end{equation}
which accordingly can be recast as follows,
\begin{equation}\label{sloweqnsderivation}
\epsilon \ll 1\, , \quad \eta \ll 1\, ,
\end{equation}
These two conditions (\ref{sloweqnsderivation}) constitute the slow-roll conditions. The parameters $\epsilon$ and $\eta$ are known as the slow-roll parameters, and these, in the context of slow-roll approximation, are formally defined to be equal to,
\begin{equation}\label{slowrolpparaenedefs}
\epsilon =\frac{1}{2\kappa^2}\left(\frac{V '(\varphi )}{V (\varphi )}\right)^2,{\,}{\,}{\,}\eta =\frac{1}{\kappa^2}\left(\frac{V''(\varphi )}{V (\varphi ^2)}\right) \, .
\end{equation}
Notice that the prime in all the equations above denotes differentiation with respect to the canonical scalar field $\varphi $. The observational indices that are currently scrutinized by the Planck collaboration \cite{planck}, are written in terms of the slow-roll parameters. Particularly we shall be interested in the spectral index of the primordial curvature fluctuations, denoted as $n_s$, and the tensor-to-scalar ratio, denoted as $r$, which are expressed in terms of the slow-roll parameters (\ref{slowrolpparaenedefs}) in the following way \cite{inflation,inflationreview},
\begin{equation}\label{observquantit}
n_s\sim 1-6\epsilon+2\eta\, , \quad r=16\epsilon .
\end{equation}
Finally, the latest Planck (2015) observational data \cite{planck} predict for the observational indices of Eq. (\ref{observquantit}), the following values,
\begin{equation}\label{planckconstr}
n_s=0.9655\pm 0.0062\, , \quad r<0.11.
\end{equation}
After this brief introduction to the slow-roll approximation and related observational indices, we now proceed to the phenomenological analysis of the Einstein frame canonical scalar field theory counterpart of the $F(R)$ gravity (\ref{jordanframegravity}), with Einstein frame canonical scalar potential (\ref{skf}). The model of Eq. (\ref{skf}) was also presented in Ref. \cite{sergeistarobinsky}, to which reference we refer the reader for further details. As it can be shown, the minimum of the canonical scalar potential (\ref{skf}), is at $\varphi=0 $, since the critical point of the equation $V'(\varphi)=0$ is actually $\varphi=0$. Also, since the second derivative of the potential at $\varphi=0$ is, 
\begin{equation}\label{secondderivpot}
V''(0)=-\frac{4 C_0 \kappa ^2}{3}+\frac{8 C_2 \kappa ^2}{3}
\end{equation}
which is positive when $C_2>\frac{C_0}{2}$, the critical point $\varphi=0$ is a global minimum of the potential $V(\varphi )$. We recall here that the inflationary evolution of the canonical scalar field $\varphi$ goes as follows: during the inflationary era, the scalar field has very large values ($\varphi$) and inflation ends when the slow-roll parameters of Eq. (\ref{slowrolpparaenedefs}), become of the order one. Eventually, the field tends to the value $\varphi$, which corresponds to the minimum of the potential. For the potential (\ref{skf}), the Eqs. (\ref{eqnsmotoinscla}) and (\ref{frweqnbecomes}) in the slow-roll approximation become,
\begin{equation}\label{neeeqasw}
\frac{3H^2}{\kappa^2}\simeq \frac{\gamma -2 e^{-\sqrt{\frac{2}{3}} \kappa  \varphi } \gamma +4 C_2 e^{-2 \sqrt{\frac{2}{3}} \kappa  \varphi } \kappa ^2}{4 \kappa ^2},{\,}{\,}{\,}3H\dot{\varphi}\simeq -\frac{e^{-2 \sqrt{\frac{2}{3}} \kappa  \varphi } \left(e^{\sqrt{\frac{2}{3}} \kappa  \varphi } \gamma -4 C_2 \kappa ^2\right)}{\sqrt{6} \kappa } 
\end{equation}
where for convenience we introduced the parameter $\gamma$, which is defined in terms of $C_0$ as follows,
\begin{equation}\label{gammapardefintrebd}
\gamma=4\kappa^2 C_0.
\end{equation}
Since during inflation, the canonical scalar field has large values ($\varphi\rightarrow \infty$), the FRW equations and the scalar field equations (\ref{neeeqasw}) become,
\begin{equation}\label{slowrollduringinflation}
H^2\simeq \frac{\gamma}{12},{\,}{\,}{\,}3H\dot{\varphi}\simeq \left(\frac{\gamma}{\sqrt{6\kappa^2}}\right)e^{-\sqrt{\frac{2}{3}}\kappa   \varphi },
\end{equation}
where we ignored subdominant terms and more importantly we assumed that,
\begin{equation}\label{gammacond}
\gamma^2\gg C_2\kappa^2,
\end{equation}
which ensures that a quasi de Sitter solution can be achieved. So during inflation, where $\varphi \rightarrow \infty$, the slow-roll indices read,
\begin{equation}\label{slowrollduring}
\epsilon=\frac{4}{3 \left(-2+e^{\sqrt{\frac{2}{3}} \kappa  \varphi }\right)^2},{\,}{\,}{\,}\eta\simeq \frac{4}{3\left |2-e^{\sqrt{\frac{2}{3}} \kappa  \varphi }\right |}.
\end{equation} 
The inflationary era ends for the value of $\varphi$, for which the parameter $\epsilon$ (or $\eta$) becomes of order one, that is, $\epsilon( \varphi_{end})\simeq 1$, which for the parameter $\epsilon$ appearing in Eq. (\ref{slowrollduring}), occurs for,
\begin{equation}\label{varphioccurs}
\varphi_{end}\simeq -\sqrt{\frac{3}{2}}\frac{1}{\kappa}\ln \left(2-\sqrt{\frac{4}{3}}\right).
\end{equation}
By using Eq. (\ref{slowrollduring}), the observational indices can be written as follows,
\begin{equation}\label{actualformofobservindi}
n_s\simeq 1+\frac{8}{6-3 e^{\sqrt{\frac{2}{3}} \kappa  \varphi }}-\frac{8}{\left(-2+e^{\sqrt{\frac{2}{3}} \kappa  \varphi }\right)^2},{\,}{\,}{\,}r\simeq  \frac{64}{3 \left(-2+e^{\sqrt{\frac{2}{3}} \kappa  \varphi }\right)^2}.
\end{equation}
We can write these by using the $e-$folding number as follows (see also \cite{sergeistarobinsky}):,
\begin{equation}\label{spectrobserv}
n_s\simeq 1-\frac{2}{N},{\,}{\,}r\simeq \frac{12}{N^2}.
\end{equation}
In order to see this, recall that the $e-$folding number in the slow-roll approximation is defined to be equal to,
\begin{equation}\label{efoldingslowrollacdef}
N\simeq \kappa^2 \int_{\varphi_{end}}^{\varphi^*}\frac{V(\varphi)}{V'(\varphi)}\mathrm{d}\varphi,
\end{equation}
where $\varphi^*$, is an initial value of the scalar field $\varphi$ which is assumed to be $\varphi^* \gg \varphi_{end}$. By using this approximation, we obtain that,
\begin{equation}\label{neacttval}
N=\frac{3}{4} e^{\sqrt{\frac{2}{3}} \varphi^* \kappa }
\end{equation}
for which value, the slow-roll parameters become,
\begin{equation}\label{slowrollduringnewones}
\epsilon=\frac{4}{3 \left(e^{\sqrt{\frac{2}{3}} \kappa  \varphi^* }\right)^2},{\,}{\,}{\,}\eta\simeq \frac{4}{3\left|e^{\sqrt{\frac{2}{3}} \kappa  \varphi^* }\right|}.
\end{equation} 
and by combining Eqs. (\ref{neacttval}) and (\ref{slowrollduringnewones}), we easily obtain Eq. (\ref{spectrobserv}). Then, by looking Eq. (\ref{spectrobserv}), in order to achieve for example $N=60$ $e-$folding, it is required that the initial value of the scalar field is approximately, $\varphi_i\simeq \frac{1.07}{\kappa^2} $ (see also \cite{sergeistarobinsky} for details). The resulting picture is compatible with the 2015 Planck data \cite{planck}, since for $N= 60$, we obtain that $n_s\simeq 0.9665$, and $r\simeq 0.0029$. In conclusion, as was also pointed out in detail in \cite{sergeistarobinsky}, the resulting picture is pretty much alike the standard $R^2$ model, with the only difference being traced on the fact that the minimum of the potential is non-zero. Therefore, the $F(R)$ gravity model of (\ref{jordanframegravity}), which describes a Jordan frame $R^2$ gravity plus cosmological constant, in the Jordan frame can give rise to the singular bounce of Eq. (\ref{scalebounce}), and when the metric in the Jordan frame is appropriately chosen, the same model can give rise to an Einstein frame inflationary potential which is a modification of the $R^2$ model, with observational indices compatible with current observational data.

\section{The effective equation of state for the singular bounce and comparison with non singular bounce}

\begin{figure}[h]
\centering
\includegraphics[width=15pc]{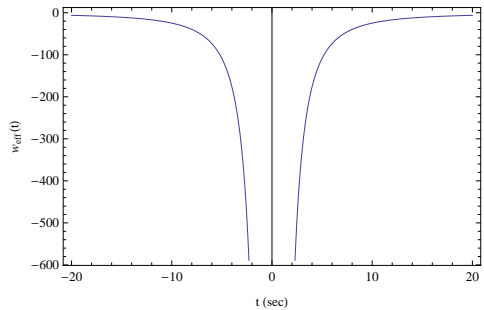}
\caption{The equation of state corresponding to the cosmological
bounce $a(t)=e^{f_0\left(t-t_s\right)^{2(1+\varepsilon)}}$, as a
function of time, for $\varepsilon=\frac{1}{11}$, $f_0=0.0001(\mathrm{sec})^{-2\varepsilon-2}$ and
$t_s=10^{-35}$sec.} \label{plot3}
\end{figure}

\begin{figure}[h]
\centering
\includegraphics[width=15pc]{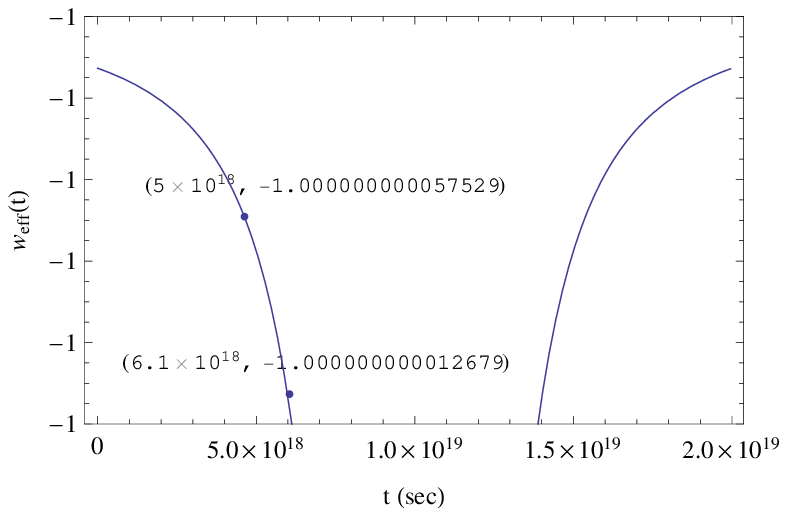}
\includegraphics[width=15pc]{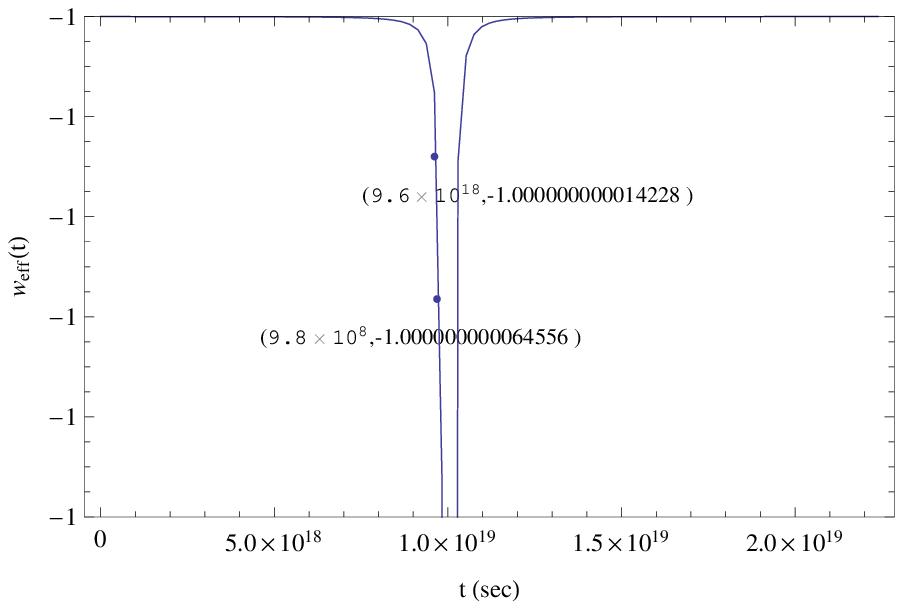}
\caption{The equation of state corresponding to the cosmological
bounce $a(t)=e^{f_0\left(t-t_s\right)^{2(1+\varepsilon)}}$, as a
function of time, for $\varepsilon=\frac{1}{11}$, $f_0=10^{-32}(\mathrm{sec})^{-2\varepsilon-2}$ (left),
$f_0=10^{-28}$ (right) and $t_s=10^{19}$sec.} \label{plot4}
\end{figure}

\begin{figure}[h]
\centering
\includegraphics[width=16pc]{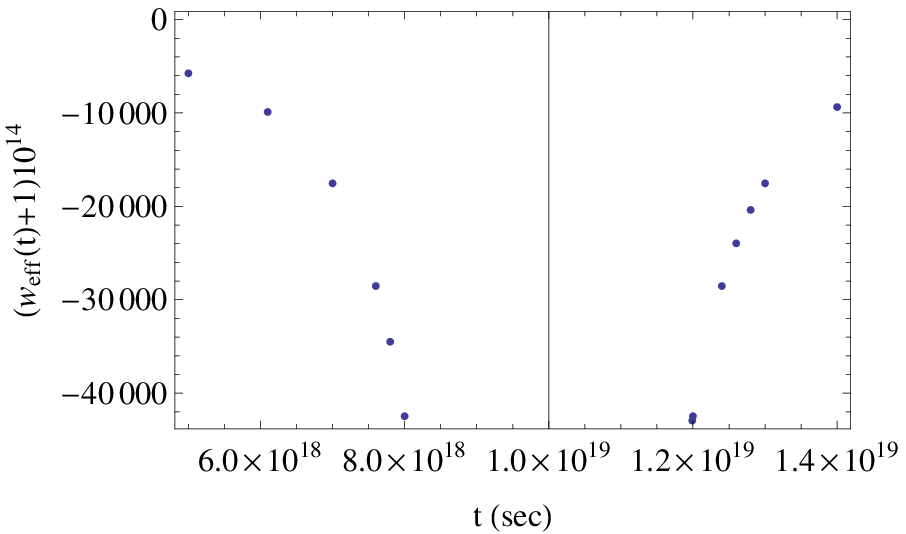}
\includegraphics[width=15pc]{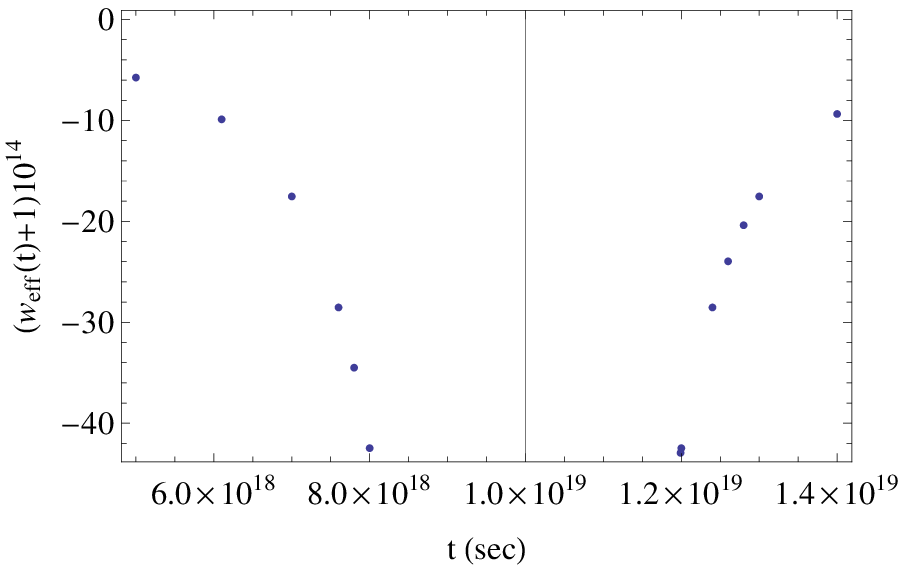}
\caption{The function $\left(w_{\mathrm{eff}}+1\right)\times 10^{14}$ corresponding to the cosmological
bounce $a(t)=e^{f_0\left(t-t_s\right)^{2(1+\varepsilon)}}$, as a
function of time, for $\varepsilon=\frac{1}{11}$, $f_0=10^{-32}(\mathrm{sec})^{-2\varepsilon-2}$ (left),
$f_0=10^{-28}(\mathrm{sec})^{-2\varepsilon-2}$ (right) and $t_s=10^{19}$sec.} \label{yog}
\end{figure}

Having studied the $F(R)$ gravity that can generate the Type IV
singularity, in this section we shall study in detail the EoS
corresponding to the singular bounce of Eq. (\ref{scalebounce}). For both a standard Einstein-Hilbert gravity background and a modified gravity background\footnote{Equation (\ref{eosdef}) is valid when the $F(R)$ geometric contribution is viewed as a perfect fluid, see Appendix A for details}, the EoS for a Hubble rate $H(t)$ is defined to be \cite{reviews},
\begin{equation}\label{eosdef}
w_{\mathrm{eff}}=-1-\frac{2\dot{H}(t)}{3H(t)^2},
\end{equation}
so for the Hubble rate of Eq. (\ref{hubblebounce}), the EoS reads,
\begin{equation}\label{eoseforhub}
w_{\mathrm{eff}}=-1-\frac{(1+2 \varepsilon )}{3 f_0
(1+\varepsilon ) (t-t_s)^{2(1+\varepsilon )} }
\end{equation}
In Fig. (\ref{plot3}) we plot the EoS parameter as a function of
time, for $\varepsilon=\frac{1}{11}$, $f_0=0.0001(\mathrm{sec})^{-2\varepsilon-2}$ and $t_s=10^{-35}$sec.
As we can see, the EoS describes a phantom evolution, for $t<t_s$,
at $t\simeq t_s$ it develops a singularity at some time during the phantom era, and
for $t>t_s$ slowly by slowly goes from the phantom era to the de
Sitter evolution. Therefore for $t\gg t_s$, and $t\ll t_s$, the
evolution is nearly de Sitter. This singular behavior in the deep
phantom era has also been observed in other bouncing models too, see
for example \cite{quintombounce}, and we shall briefly discuss this
issue later on.
We can also study the behavior of the EoS analytically in various
limits of the cosmic time. Let us start with the case that the
bouncing point, and simultaneously the point that the Type IV
singularity occurs, is very small, that is $t_s/t\ll 1$. In such a
case, for large $t$ values, and specifically for all cosmic times
that $t\gg t_s$, the EoS would become,
\begin{equation}\label{eoseforhub1}
w_{\mathrm{eff}}=-1-\frac{(1+2 \varepsilon )}{3 f_0
(1+\varepsilon ) (t)^{2(1+\varepsilon )} },
\end{equation}
which for large $t$ describes de Sitter acceleration that slightly
crosses the phantom divide. Moreover, for times near
the Type IV singularity, that is $t\simeq t_s$, the EoS becomes
singular, as it can be seen from Eq. (\ref{eoseforhub}). This can
also be seen in Fig. \ref{plot3}. Now consider the case that $t_s\gg
t$, in which case, when $t\simeq t_s$, the same singular behavior
occurs and the EoS evolves rapidly in the deep phantom era. For
cosmic times much smaller than $t_s$, that is $t_s\gg t$, the EoS is
approximately equal to,
\begin{equation}\label{eoseforhub2}
w_{\mathrm{eff}}=-1-\frac{ (1+2 \varepsilon )}{3 f_0
(1+\varepsilon ) t_s^{2(1+\varepsilon )}},
\end{equation}
which again describes nearly de Sitter but slightly crosses to the
phantom state. It is worth examining another interesting scenario,
in which the EoS evolves to the singular phantom state after our
present epoch, which is roughly $t_p\simeq 10^{17}$sec. In that case
the EoS up to our era is nearly de Sitter, and evolves to phantom
near the Type IV singularity. For example let $t_s=10^{19}$sec.
Then, for $t<t_s$, the EoS is of nearly de Sitter type and as $t\to
t_s$, the EoS becomes rapidly phantom. This can also be seen in Fig.
\ref{plot4}, left panel, where we used $f_0=10^{-28}(\mathrm{sec})^{-2\varepsilon-2}$. In the right
panel we have plotted the EoS for $f_0=10^{-15}(\mathrm{sec})^{-2\varepsilon-2}$. Observe that the
``throat'' of the EoS graph becomes smaller, as the value of $f_0$
increases.
So practically, according to the left plot scenario of Fig.
\ref{plot4}, this scenario suggests that the EoS can
possibly and simultaneously cross the phantom divide, with a functional behavior that resembles to a Dirac function. This seems somehow like a simultaneous and peculiarly
sudden change in the state of the Universe as a whole. It would be
interesting to model such a behavior with a Dirac delta function at
the transition, to see the differences between the bounce model
(\ref{scalebounce}) and the Dirac delta function model, but we defer this task to a future work. In the plots of Fig. \ref{plot4}, there appear many "$-1$'', and this is due to the fact that the values of the EoS are very close to $-1$, with the differences appearing after 12 decimal places. In order to have a more clear picture of this, in Table \ref{tableeos}, we present the exact values of the EoS for various times and for $f_0=10^{-32}(\mathrm{sec})^{-2\varepsilon-2}$, where the differences between various points can explicitly be seen.
\begin{table*}
\small
\caption{\label{tableeos}Exact Values of the EoS for $f_0=10^{-32}(\mathrm{sec})^{-2\varepsilon-2}$}
\begin{tabular}{@{}crrrrrrrrrrr@{}}
\tableline
 $t$ (sec) & $w_{\mathrm{eff}}$\,\,\,\,  
\\
\tableline
 $t=5\times 10^{18}$ & $w_{\mathrm{eff}}=-1.000000000000058$ \\
\tableline
 $t=6.1\times 10^{18}$& $w_{\mathrm{eff}}=-1.000000000000099$ \\
\tableline 
$t=7\times 10^{18}$& $w_{\mathrm{eff}}=-1.000000000000175$  \\
\tableline
$t=8\times 10^{18}$& $w_{\mathrm{eff}}=-1.000000000000425$  \\
\tableline
 $t=1.1\times 10^{19}$& $w_{\mathrm{eff}}=-1.000000000001927$ \\
\tableline
 $t=1.5\times 10^{19}$& $w_{\mathrm{eff}}=-1.000000000000058$ \\
\tableline
 \end{tabular}
\end{table*}
Furthermore, in Table \ref{tableos} we present the corresponding exact EoS values for $f_0=10^{-28}(\mathrm{sec})^{-2\varepsilon-2}$.
\begin{table*}
\small
\caption{\label{tableos}Exact Values of the EoS for $f_0=10^{-32}(\mathrm{sec})^{-2\varepsilon-2}$}
\begin{tabular}{@{}crrrrrrrrrrr@{}}
\tableline
 $t$ (sec) & $w_{\mathrm{eff}}$\,\,\,\,  
\\
\tableline
 $t=5\times 10^{18}$ & $w_{\mathrm{eff}}=-1.000000000057529$ \\
\tableline
 $t=6.1\times 10^{18}$& $w_{\mathrm{eff}}=-1.000000000012679$ \\
\tableline 
$t=7\times 10^{18}$& $w_{\mathrm{eff}}=-1.000000000175356$  \\
\tableline
$t=8\times 10^{18}$& $w_{\mathrm{eff}}=-1.000000000424738$  \\
\tableline
 $t=1.1\times 10^{19}$& $w_{\mathrm{eff}}=-1.000000001927141$ \\
\tableline
 $t=1.2\times 10^{19}$& $w_{\mathrm{eff}}=-1.000000000424738$ \\
\tableline
 $t=1.5\times 10^{19}$& $w_{\mathrm{eff}}=-1.000000000057529$ \\
\tableline
 \end{tabular}
\end{table*}
In order to have a clear picture of the behavior of the EoS, in Fig. \ref{yog} we plot the behavior of the function $\left(w_{\mathrm{eff}}+1\right)\times 10^{14}$ as a function of the cosmic time for $f_0=10^{-32}(\mathrm{sec})^{-2\varepsilon-2}$ (left) and $f_0=10^{-28}(\mathrm{sec})^{-2\varepsilon-2}$ (right). In these plots, the differences of the EoS in the two cases can be clearly seen. The vertical line in the plots of Fig. \ref{yog}, corresponds to the value $t=10^{19}$sec.

\begin{figure}[h]
\centering
\includegraphics[width=15pc]{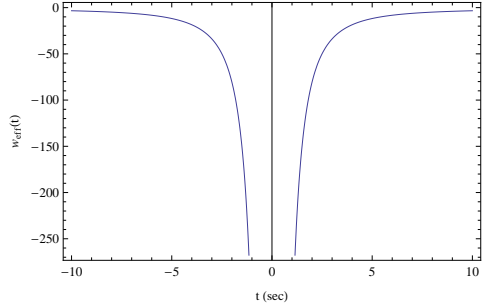}
\caption{The equation of state corresponding to the cosmological
bounce $a(t)=e^{f_0\left(t-t_s\right)^{2}}$, as a function of time,
for $f_0=0.0001(\mathrm{sec})^{-2\varepsilon-2}$ and $t_s=0$sec.} \label{plot5}
\end{figure}

Let us now in brief compare the behavior of the EoS corresponding to
the singular bounce (\ref{scalebounce}) to the one corresponding to
the bounce (\ref{bambabounce}). The two models as we now explicitly
demonstrate are qualitatively similar, with the only difference
being that the two models are generated by different $F(R)$
gravities in the Jordan frame. For an account on the bounce
(\ref{bambabounce}), see \cite{sergeibabmba}. For the bounce
(\ref{bambabounce}), the EoS reads,
\begin{equation}\label{eosbma}
w_{\mathrm{eff}}=-1-\frac{1}{3 f_0 (t-t_s)^2}
\end{equation}
In Fig. \ref{plot5} we plotted the time dependence of the bounce
(\ref{bambabounce}), for $f_0=0.0001(\mathrm{sec})^{-2\varepsilon-2}$ and $t_s=0$.
As is obvious from Fig. \ref{plot5}, and as we already mentioned, the
qualitative behavior of the bounces (\ref{scalebounce}) and
(\ref{bambabounce}), are very much alike. It is of particular
importance to notice the appearance of a singularity in the EoS of
the two bounces and also that both bounces in the limit $t\gg t_s$
asymptotically approach a de Sitter evolution. Moreover, near the
bouncing point the two bounces generate a phantom EoS. Notice
however that the bounce of Eq. (\ref{bambabounce}) is not singular
at all (the derivative of the Hubble rate is constant and all higher
derivatives are simply zero). But what does a singularity in the
equation of state indicates? In the next section we shall briefly
discuss this issue and by using some illustrative examples, always
in the context of bouncing cosmologies, we try to shed some light on
this issue.

\subsection{Brief discussion on singularities in the effective
equation of state in general bouncing cosmologies}

\begin{figure}[h] \centering
\includegraphics[width=15pc]{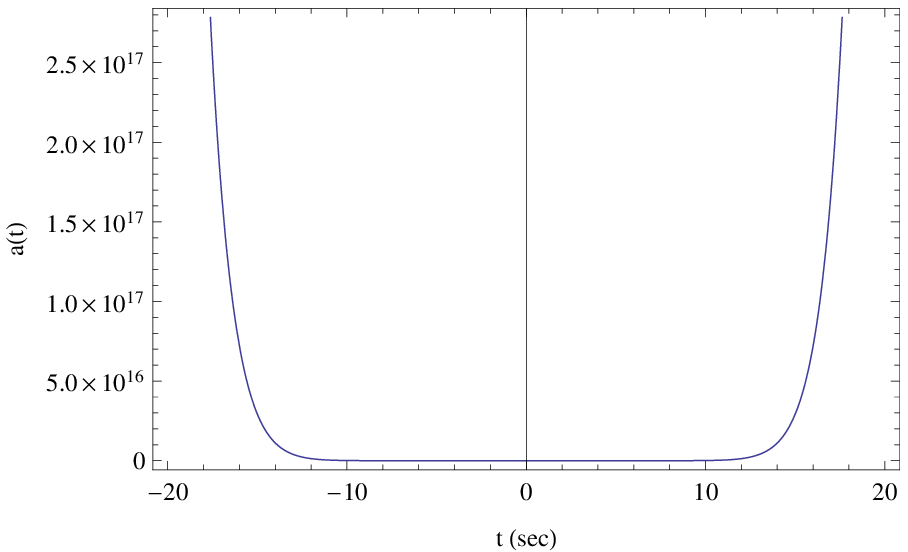}
\includegraphics[width=15pc]{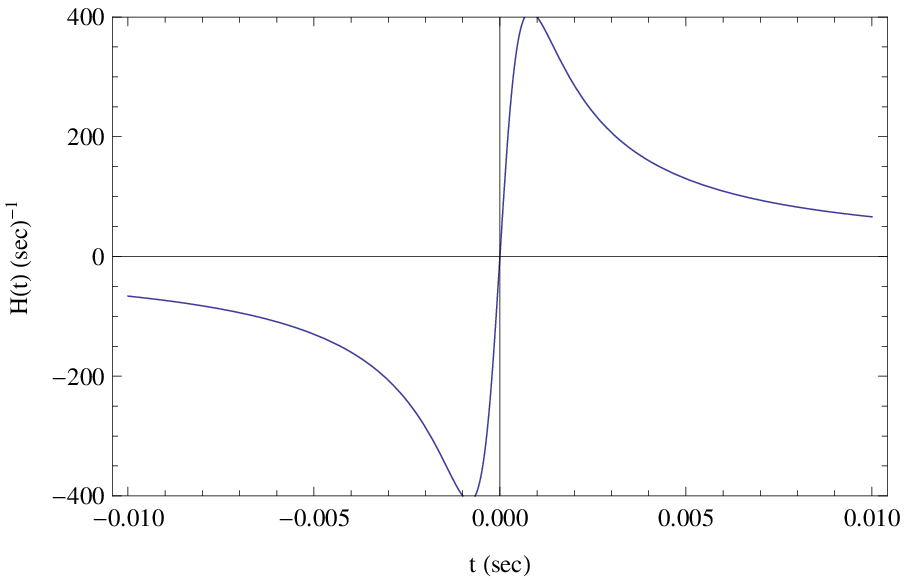}
\caption{The scale factor $a(t)$ (left plot) and the Hubble rate
(right plot) as a function of the cosmic time $t$, for
$t_s=10^{-35}$sec, and $\rho_c=10^6\mathrm{J}/m^3$ for $a(t)=\left(\frac{3}{2} \rho_c t^2+1\right)^{\frac{1}{3}}$}
\label{superplot}
\end{figure}

In this section we shall discuss the issue of singularities of the EoS in the context of bouncing cosmology. Apart from the singular bounce which we studied in this paper, there exist also other bouncing cosmologies, such as the superbounce \cite{bounce4,superbounce2,superbounce3}, and the matter bounce \cite{matterbounce} scenarios. It is worth discussing briefly how the EoS behaves in these bouncing cosmologies. We start off with the matter bounce scenario \cite{matterbounce}, for which the scale factor reads,
\begin{equation}\label{matterbouncescale}
a(t)=\left(\frac{3}{2} \rho_c t^2+1\right)^{\frac{1}{3}},
\end{equation}
and the corresponding Hubble rate is equal to,
\begin{equation}\label{hubratematterb}
H(t)=\frac{2 t \rho_c}{2+3 t^2 \rho_c}.
\end{equation}
where $\rho_c$ is the critical energy density. This parameter frequently occurs in the LQC matter bounce scenario, and it can take various values depending on the model under study (see for example page 17 of \cite{matterbounceoik}, for a detailed discussion).
The matter bounce cosmology described by the Eqs. (\ref{matterbouncescale}) and (\ref{hubratematterb}) has all the bouncing cosmology features we described earlier, as can also be easily verified by looking Fig. (\ref{superplot}), where we plotted the time dependence of the scale factor $a(t)$ and of the Hubble rate $H(t)$, as a function of time, for $\rho_c=10^6\mathrm{J}/\mathrm{m}^3$.
Of course, in the matter bounce case, the bouncing point is $t=0$. The EoS corresponding to matter bounce scenario reads,
\begin{equation}\label{eosemattebounce}
w_{\mathrm{eff}}=-1-\frac{2 \dot{H}(t)}{3 H(t)^2}=-\frac{2}{3 t^2 \rho_c}.
\end{equation}
Clearly, the EoS (\ref{eosemattebounce}) is singular at the bouncing point $t=0$. The case of the matter bounce scenario is very similar to the singular bounce case we studied in this paper, with scale factor given in Eq. (\ref{scalebounce}) and also to the bounce of Eq. (\ref{bambabounce}) studied in \cite{sergeibabmba}. In addition, the quintom bounce studied in Ref. \cite{quintombounce} has exactly the same characteristics as the singular bounce we studied in this article and also as the matter bounce scenario has, although in \cite{quintombounce}, standard Einstein Hilbert gravity was employed, with regards to the $F(R)$ content of the theory. Notice that in all three aforementioned cases, the EoS is singular at the bouncing point, which seems to be a repeating pattern, common in certain types of bouncing cosmologies. We need to stress that a singular EoS in astrophysical systems is a rather peculiar feature which never occurs, so in principle, the same would be expected in cosmological theories. So what we need to understand is what does that singularity in the EoS indicates and how it is related to the finite time singularities we presented in a previous section. Also we have to note that in both the matter bounce and singular bounce case, the Hubble rate vanishes at the bouncing point, so probably a singularity in the EoS is unavoidable, unless this is somehow cancelled if $H^2$ behaves in the same way as $\dot{H}$. However, for the functional forms of the Hubble rate in matter bounce and the singular bounce cases, this requirement is not fulfilled. In addition, note that the fact that EoS parameter $w_{\mathrm{eff}}$ is singular does not always imply that the effective pressure and/or effective energy-density are singular too \cite{Nojiri:2005sx,babrowski}. Both of them maybe regular while the effective pressure and effective energy density might be finite.

For example, there exist some other bouncing cosmologies in the literature, such as the superbounce \cite{bounce4,superbounce2,superbounce3}, for which the requirement that $H^2$ behaves in the same way as $\dot{H}$ occurs. In this case, no singularity occurs in the EoS. It is worth recalling this case in brief for completeness. In the superbounce scenario \cite{bounce4,superbounce2,superbounce3}, the scale factor is equal to,
\begin{equation}\label{superbouncescale}
a(t)=\left(t-t_s\right)^{\frac{2}{c^2}},
\end{equation}
and the corresponding Hubble rate is equal to,
\begin{equation}\label{hubratesuperb}
H(t)=\frac{2}{c^2 (t-t_s)}.
\end{equation}
where $c$ is a dimensionless parameter, which is assumed to satisfy $c>\sqrt{6}$ in order for the superbounce to occur(see \cite{superbounce3} and references therein). The superbounce (\ref{superbouncescale}) clearly describes a bouncing cosmology, having most of the properties of a bounce we described in a previous section, with the only exception being that at the bouncing point $t=t_s$, the Hubble rate is not zero but diverges. This can also be verified by looking Fig. (\ref{superplot}), where we plotted the time dependence of the scale factor and of the Hubble rate, for $c=\sqrt{7}$ and $t_s=10^{-35}$sec. 
\begin{figure}[h] \centering
\includegraphics[width=15pc]{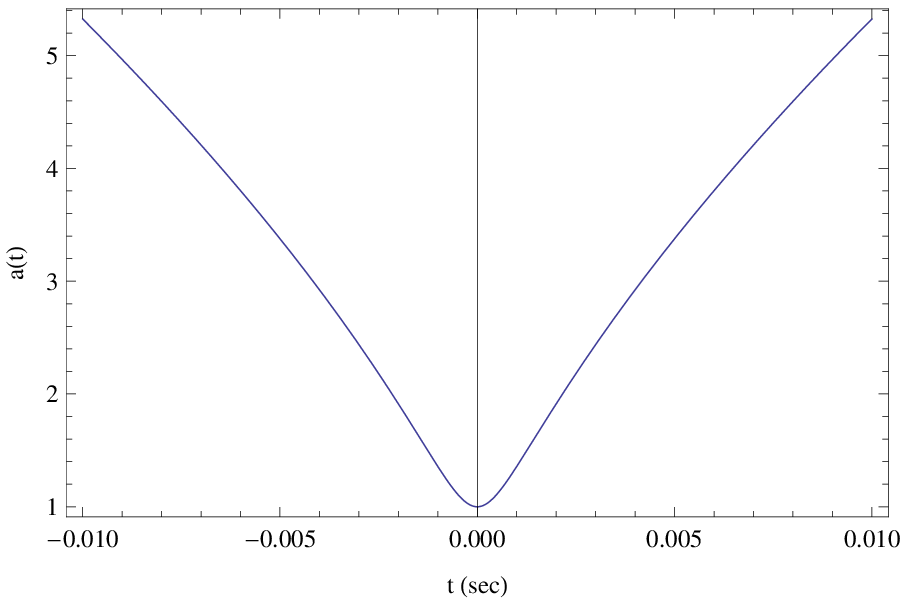}
\includegraphics[width=15pc]{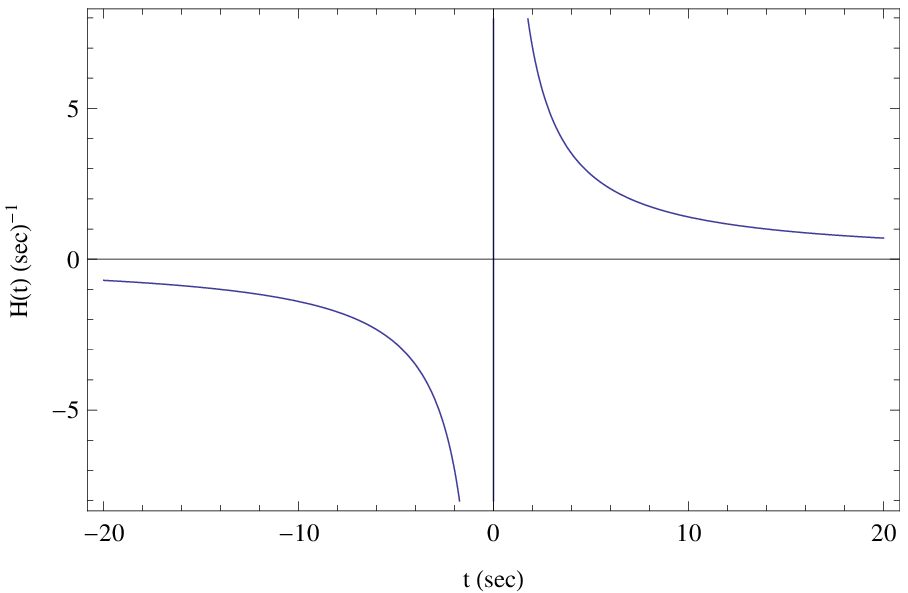}
\caption{The scale factor $a(t)$ (left plot) and the Hubble rate
(right plot) as a function of the cosmic time $t$, for
$t_s=10^{-35}$sec, and $c=\sqrt{7}$ for $a(t)=\left(t-t_s\right)^{\frac{2}{c^2}}$}
\label{superplot1}
\end{figure}
As we can see in Fig. (\ref{superplot1}), the scale factor is never singular and the Universe contracts for $t<t_s$ and expands for $t>t_s$, where at $t=t_s$ no singularity occurs.  In addition, the EoS for the superbounce (\ref{superbouncescale}) is equal to,
\begin{equation}\label{eosesupebounce}
w_{\mathrm{eff}}=\frac{1}{3} \left(-3+c^2\right),
\end{equation}
which is constant. So in the superbounce case, only the Hubble rate is singular, with the EoS being non-singular and constant and this occurs because $H^2$ and $\dot{H}$ behave in the same way as functions of the cosmic time. This is in contrast to the singular bounce we studied in this paper (\ref{scalebounce}), the bounce (\ref{bambabounce}), which was studied in \cite{sergeibabmba} and also to the quintom bounce which was studied in \cite{quintombounce}. Notice that in all the three aforementioned cases, at the bouncing point $t=t_s$, the Hubble rate vanishes, that is, $H(t_s)=0$. Someone could claim that for example, the singularity in the EoS, is due to the reason that $H(t_s)=0$, which is true in both cases of the bounces. So provisionally, we could claim that the singularity in the EoS is, at a first glance, not directly related to the finite time cosmological singularities in the case that the Type IV singularity is taken into account and only that. This however should be studied in more detail and caution and therefore, the need for the complete understanding of the relation between the finite time singularities and the EoS singularities, is compelling. This however exceeds the purposes of this paper and we hope to address this issue concretely in a future work.

\section{Conclusions}

In this paper we investigated which $F(R)$ gravity can generate a Type IV singular bounce, which was chosen so that the Type IV singularity occurs at the bouncing point. As we explicitly demonstrated, the $F(R)$ gravity responsible for the Type IV bounce, near the singularity has the form $F(R)=R+\alpha R^2+\Lambda$ in the Jordan frame. We also found the Einstein frame scalar theory counterpart corresponding to the $F(R)$ gravity and also, having assumed a quasi de Sitter metric in the Einstein frame, we investigated the inflationary properties of the resulting scalar theory in the Einstein frame. As we evinced, the spectral indices of the resulting scalar theory can be compatible with the recent Planck \cite{planck} observations and the theory it self is compatible with the standard Einstein frame $R^2$ inflation theory, with the only difference being that the minimum of the potential is shifted.

We also performed a thorough analysis of the EoS corresponding to the Type IV singular bounce. As we showed, the behavior of the singular bounce is similar to other bounces that exist in the literature, such as the matter bounce scenario \cite{matterbounce} or the quintom bounce scenario \cite{quintombounce}. At this point it is worth discussing a quite interesting point that resulted from our analysis. As we demonstrated, even in the case of the Type IV singular bounce, the EoS is singular at the bouncing point, where the singularity occurs, with the other two aforementioned bounces also having this feature. In addition, as can be easily checked, in the case of a Type I singularity, the EoS can in some cases be singular or regular at the point where the singularity occurs and the same applies for the Type II case. Furthermore, it has pointed out in the literature that there exist points at which the EoS diverges but both the effective energy density and the effective pressure are finite \cite{babrowski}, so this would, in some sense, have nothing to do with finite time singularities. As for the Type II case, the EoS is always singular. It would therefore be interesting to address the question, whether the finite time singularities should be further classified by also taking into account the EoS. This is also motivated by the fact that in most known perfect fluids that are used in astrophysics, the EoS is never singular, so the appearance of this singular behavior in some cases, in a cosmological context should be noticed and used in some classifying way. This is also supported by the fact that, even in the Big Rip case, which is a crushing type singularity, the EoS can be regular, even though all other quantities diverge. In any case, the singularity in the EoS of perfect fluids used in cosmology should be further studied in order to understand the implications this would have at a theoretical and at an observational level. For a relevant study on these issues, see \cite{babrowski}.

Another important issue that we did not address is the connection between the Type IV singularities and bouncing cosmology. Is it possible that a Type IV singular cosmological evolution would lead to a bouncing cosmology? And in addition, is the EoS always singular in the case of a Type IV singular cosmological evolution? We hope to address these issues formally in a future work.

\section*{Acknowledgments}

The work of S.D.O. is supported by MINECO (Spain), projects FIS2010-15640
and FIS2013-44881.

\section*{Appendix A:EoS for $F(R)$ modified gravity}

In this appendix, we demonstrate in detail that the effective equation of state of the $F(R)$ modified gravity, for a flat FRW background, is given by Eq. (\ref{eosdef}), namely,
\begin{equation}\label{eosdefappendixversion}
w_{\mathrm{eff}}=-1-\frac{2\dot{H}(t)}{3H(t)^2}.
\end{equation}
For a detailed account on these issues, see also \cite{reviews} and references therein. We start off with the FRW equations of motion, which can be derived from Eq. (\ref{modifiedeinsteineqns}), and these read,
\begin{align}\label{flrw}
& 3F'(R)H^2=\kappa^2(\rho_m+\rho_r)+\frac{(F'(R)R-F(R))}{2}-3H\dot{F}'(R), \\ \notag &
-2F'(R)\dot{H}=\kappa^2(p_m+4/3\rho_r)+F\ddot{F}'(R)-H\dot{F}'(R),
\end{align}
so by assuming that no matter fluids are present, the above equations are modified as follows,
\begin{align}\label{flrw1}
& 3F'(R)H^2=\frac{(F'(R)R-F(R))}{2}-3H\dot{F}'(R), \\ \notag &
-2F'(R)\dot{H}=F\ddot{F}'(R)-H\dot{F}'(R),
\end{align}
The equations (\ref{flrw1}) can be rewritten as follows,
\begin{align}\label{flrw12}
& 3 H^2=\rho_{DE},{\,}{\,}{\,}-2\dot{H}=\rho_{DE}+P_{DE},
\end{align}
where we defined the geometric energy density $\rho_{DE}$ and the geometric effective pressure $P_{DE}$, as follows,
\begin{align}\label{matrix}
& \rho_{DE}=\frac{(F'(R)R-F(R))}{2}-3H\dot{F}'(R)+3H^2\\ \notag &
P_{DE}=F\ddot{F}'(R)+2H\dot{F}'(R)-\frac{(F'(R)R-F(R))}{2}+3\left(3H^2+2\dot{H}\right)F'(R)
\end{align}
In this way, the contribution of the $F(R)$ gravity can be viewed as a perfect fluid's one and the representation of the FRW equations appearing in Eq. (\ref{frwf1}) can be viewed as a mathematically equivalent perfect fluid representation of the ones appearing in Eq. (\ref{flrw}). This is the reason why the contribution of $F(R)$ gravity to the FRW equations is said to be owing to the existence of a geometric dark fluid. The geometric effective energy density $\rho_{DE}$ and the geometric effective pressure density satisfy the continuity equation,
\begin{equation}\label{continuity}
\dot{\rho_{DE}}+3H\left(\rho_{DE}+P_{DE}\right)=0,
\end{equation} 
The corresponding effective equation of state for the dark fluid is defined to be equal to,
\begin{equation}\label{darkfluidweff}
w_{\mathrm{eff}}=\frac{P_{DE}}{\rho_{DE}},
\end{equation}
which owing to Eq. (\ref{flrw1}) can be written as follows,
\begin{equation}\label{eosdefappendixversionnew}
w_{\mathrm{eff}}=-\frac{2\dot{H}+3H(t)^2}{3H^2},
\end{equation}
which can easily be cast in the form of Eq. (\ref{eosdefappendixversion}). So for a flat FRW metric, the modified gravity effective equation of state can take the form of Eq. (\ref{eosdefappendixversion}), like in the ordinary Einstein-Hilbert gravity in a flat FRW background.

\section*{Appendix B: Analytic form of constants}

Here we present the full form of the constants appearing in Eqs. (\ref{pseries}) and (\ref{qsapprox}). The parameter $\mathcal{A}$ is equal to,
\begin{equation}
\label{parametalpha} \mathcal{A}=\frac{f_0^{-\frac{1}{2 (1+\varepsilon )}}(2+2 \varepsilon )^{\frac{1}{2 (1+\varepsilon )}} C_1\Gamma\left(\frac{1}{2 (1+\varepsilon )}\right)}{2 (1+\varepsilon ) \Gamma\left(-\frac{1+4 \varepsilon }{2 (1+\varepsilon )}\right)}+\frac{f_0^{-\frac{1}{2 (1+\varepsilon )}} C_1(2+2 \varepsilon )^{\frac{1}{2 (1+\varepsilon )}} \left(1+5 \varepsilon +2 \varepsilon ^2\right) \Gamma\left(1+\frac{1}{2 (1+\varepsilon )}\right)}{2 (1+\varepsilon )^2 \Gamma\left(1-\frac{1+4 \varepsilon }{2 (1+\varepsilon )}\right)}\, .
\end{equation}
while the parameters $\mathcal{B}$ and $\mathcal{C}$ are equal to,
\begin{align}
\label{mathbandc} \mathcal{B}=&6 f_0 (1+4\varepsilon ) C_1
\Big{(}\frac{f_0}{2(\varepsilon+1) }\Big{)}
{}^{-\frac{1}{2(\varepsilon+1) }} \\ \notag & \times
\frac{\Gamma\Big{(}1+\frac{1}{2(\varepsilon+1) }\Big{)}
\Big{(}(2+2(\varepsilon+1) ) -2 (1+ 2(\varepsilon+1) )
\Gamma\Big{(}\frac{1-2\varepsilon }{2(\varepsilon+1)
}\Big{)} \Big{)} }{ (2(\varepsilon+1) )^2 \Gamma\Big{(}\frac{1-2\varepsilon }{2(\varepsilon+1) }\Big{)}} \, , \nn \mathcal{C}=& \frac{6
f_0^{\frac{22(\varepsilon+1)+1}{2(\varepsilon+1)+1}} C_1
}{(2(\varepsilon+1) )^{3+2(\varepsilon+1)}} \\ \notag & \times
\Big{(}\frac{(2(\varepsilon+1) ) (2+2(\varepsilon+1) )
 \Gamma\Big{(}\frac{1}{2(\varepsilon+1) }\Big{)}}{\Gamma\Big{(}\frac{1+4\varepsilon }{2(\varepsilon+1) }\Big{)}}-\frac{2(\varepsilon+1)  (1+4\varepsilon )^2
  \Gamma\Big{(}1+\frac{1}{2(\varepsilon+1) }\Big{)}}{\Gamma\Big{(}\frac{1-2\varepsilon }{2(\varepsilon+1) }\Big{)}}
  \\ \notag & -\frac{4 \Big{(}1+ 2(\varepsilon+1) +2(\varepsilon+1) ^2\Big{)}
   \Gamma\Big{(}1+\frac{1}{2(\varepsilon+1) }\Big{)}}{\Gamma\Big{(}\frac{1-2\varepsilon }{2(\varepsilon+1) }\Big{)}}+\frac{4 (1+2(\varepsilon+1) )^2 (1+4\varepsilon )
   \Gamma\Big{(}2+\frac{1}{2(\varepsilon+1) }\Big{)}}{(1+2(\varepsilon+1) )
    }\Big{)} \, .
\end{align}

\end{document}